\documentclass{article}

\usepackage{arxiv}

\usepackage[utf8]{inputenc} 
\usepackage[T1]{fontenc}    
\usepackage{hyperref}       
\usepackage{url}            
\usepackage{booktabs}       
\usepackage{amsfonts}       
\usepackage{nicefrac}       
\usepackage{microtype}      
\usepackage{lipsum}		
\usepackage{graphicx}
\usepackage{natbib}
\usepackage{doi}

\usepackage{tikz}

\usepackage{amsthm}
\newtheorem{thm1}{Theorem}

\usepackage{amsmath,amsfonts,amssymb,bm}
\usepackage{mathptmx}
\usepackage{newtxtext}
\usepackage{newtxmath}
\usepackage{relsize}

\usepackage{appendix}

\usetikzlibrary{calc,trees,positioning,arrows,chains,shapes.geometric,%
    decorations.pathreplacing,decorations.pathmorphing,shapes,%
    matrix,shapes.symbols}
\tikzset{
>=stealth',
  punktchain/.style={
    rectangle, 
    rounded corners, 
    draw=black, very thick,
    text width=10em, 
    minimum height=3em, 
    minimum width=4cm,
    text centered, 
    on chain},
  line/.style={draw, thick, <-},
  element/.style={
    tape,
    top color=white,
    bottom color=blue!50!black!60!,
    minimum width=8em,
    draw=blue!40!black!90, very thick,
    text width=10em, 
    minimum height=3.5em, 
    text centered, 
    on chain},
  every join/.style={->, thick,shorten >=1pt},
  decoration={brace},
  tuborg/.style={decorate},
  tubnode/.style={midway, right=2pt},
}

\title{Stochastic Downscaling to Chaotic Weather Regimes using Spatially Conditioned Gaussian Random Fields with Adaptive Covariance}

\author{Rachel Prudden\thanks{\texttt{rachel.prudden@informaticslab.co.uk}; also affiliated with University of Exeter} \\
	Informatics Lab\\
	Met Office\\
	Exeter\\
	\\
	\And
	Niall Robinson\\
	Met Office\\
	Exeter \\
	\AND
	Peter Challenor \\
	University of Exeter \\
	Exeter \\
	\And
	Richard Everson \\
	University of Exeter \\
	Exeter \\
}

\begin{document}

\maketitle

\begin{abstract}
Downscaling aims to link the behaviour of the atmosphere at fine scales to properties measurable at coarser scales, and has the potential to provide high resolution information at a lower computational and storage cost than numerical simulation alone. This is especially appealing for targeting convective scales, which are at the edge of what is possible to simulate operationally. Since convective scale weather has a high degree of independence from larger scales, a generative approach is essential. We here propose a statistical method for downscaling moist variables to convective scales using conditional Gaussian random fields, with an application to wet bulb potential temperature (WBPT) data over the UK. Our model uses an adaptive covariance estimation to capture the variable spatial properties at convective scales. We further propose a method for the validation, which has historically been a challenge for generative models.
\end{abstract}

\makeatletter{\renewcommand*{\@makefnmark}{}
\footnotetext{\\ This work has been submitted to Weather and Forecasting. Copyright in this work may be transferred without further notice.}\makeatother}


\section{Introduction}

Accurate numerical simulation of convective-scale weather is intensive in its demand for computational resources. This is partly a consequence of increasing spatiotemporal resolution, which increases computational demand geometrically. Beyond this, the atmosphere has a high level of inherent chaotic variability at convective scales, leading to a demand for larger ensembles to capture the predictive distribution to the fullest extent possible. The challenge of meeting these competing demands for computing power, as well as the difficulty of utilising the vast amounts of data produced, lead to the question: can more be done with lower-resolution data?

Downscaling aims to address this question by linking the behaviour of the atmosphere at fine scales to properties at coarser scales. In particular, statistical downscaling defines these connections without the use of a physics-based dynamical model. Statistical downscaling thus has the potential to provide high resolution information at a lower computational and storage cost than numerical simulation alone.

In some instances small scale structure is deterministically and directly related to the large scales, for instance, when the structure is driven by orographic forcing.
In such cases, downscaling can be treated as a deterministic regression problem. The situation is similar when the target variables consist of temporal averages, as is usually the case in climate applications, because the temporal averaging has the effect of smoothing out small-scale variability. 

In other cases, notably the generation of chaotic structure, this approach becomes ineffective.
However, there is reason for optimism, as bulk statistical properties of the target field may still be strongly constrained by the larger scales: the spatial characteristics of the target field may be highly predictable even if its point values are not. In such cases, rather than making a deterministic prediction, downscaling is best viewed probabilistically. The aim is then to determine the distribution of possible target fields and draw samples from this distribution.

An example of where this approach can be valuable is in downscaling to convective scales. While convective scale weather has a high degree of independence from larger scales, its overall spatial properties tend to be more predictable. Furthermore, the spatial properties of cloud and rain are highly relevant for applications such as energy forecasting and hydrology, where spatiotemporal variability can be as important as the expected value \citep{liu_ensemble_2019, schaake_hepex:_2007}. This work directly targets stochastic downscaling of moist variables to convective scales, using a statistical model called Gaussian random fields (GRFs). GRFs are statistical models of spatial functions; like Gaussian processes \citep{Rasmussen2006Gaussian} they define a distribution over such functions with a single sample corresponding to a spatially coherent function. This spatial interpretation, combined with convenient methods for conditioning on observed data, makes them a useful tool for statistical downscaling.

In contrast with most of the GRF literature, we use spatial conditioning to model the connection between small and large scales: that is, we model our low resolution data as an areal average over the high resolution grid where the target data are defined. A more standard approach would be to model all observations as belonging to point locations, as in \cite{wikle_spatiotemporal_2001}, but this would fail to account for the ensemble properties of moist variables associated with convection at the sub-grid scale \citep{arakawa_interaction_1974}. Spatial conditioning enables a faithful representation of both scales and the relationship between them \citep{Gotway2002, Kyriakidis2004}.

Historically, the lack of a standardised and comprehensive approach to verification has been a challenge for generative models, including their application to downscaling \citep{Mathieu2015DeepError}. Standard point-based scores such as mean-squared error (MSE) are not well suited to assessing the skill of stochastic models due to issues like the double penalty problem \citep{gilleland_intercomparison_2009}. This is a known challenge in generative modelling \citep{Mathieu2015DeepError}, with the consequence that subjective value judgements and other heuristic scores are common \citep{barratt_note_2018}. In this work we take a multifaceted approach to verification, and propose a novel spatial verification metric which allows for simple comparisons between stochastic models (see also references in Section \ref{nwass}).

The main contributions of this paper are:
\begin{itemize}
    \item we explain the spatial conditioning of GRFs with reference to stochastic downscaling of atmospheric variables
    \item we introduce an adaptive GRF model for downscaling, which combines spatial conditioning with an instantaneous lengthscale estimation, allowing for differing weather regimes
    \item we introduce a novel verification score for stochastic models, the neighbourhood Wasserstein score.
\end{itemize}
We begin by describing related work in Section \ref{lit}. The problem setup is outlined in Section \ref{setup}, the theory behind our GRF model in Section \ref{method}, our approach to verification in Section \ref{verification}, and Section \ref{app} describes an application to wet bulb potential temperature over the UK.

\section{Related work} \label{lit}

\subsection{Deterministic downscaling}

Deterministic methods remain widely-used in downscaling, especially for temporally aggregated and non-moist variables. Downscaling can be treated as a determisitic regression problem by expressing the values of a high resolution weather field in terms of low resolution covariates
$$\mathbf{y} = f(\mathbf{X})$$
where $f$ is referred to as a regression or transfer function. As discussed in \cite{wilby_guidelines_2004}, the philosophy underlying these methods is that the local weather or climate scenario depends on a combination of the large-scale situation and local geographic features. From this perspective, the function of $f$ is to encode the effects of these local features.

A diverse range of algorithms may be used for the transfer function. For example, linear and ridge regression were used for site-specific downscaling of temperature and precipitation in \cite{hessami_automated_2008}, while \cite{ghosh_svm-pgsl_2010} uses support vector regression for a similar use-case. More recently, \cite{vandal_deepsd:_2017} has applied convolutional neural networks to gridded precipitation downscaling. Simple methods are often effective: BCSD \citep{wood_long-range_2002}, a linear method based on single shifting and scaling factors, was found by \cite{vandal_intercomparison_2017} to outperform more complex, non-linear approaches.

While deterministic methods can be effective for some variables, due to their underlying assumptions they are ill-suited for modelling moist variables over short time periods \citep{Raut2019, schoof2001downscaling}. This has motivated research into stochastic methods, outlined in the next section.

\subsection{Stochastic downscaling}

Stochastic downscaling using statistical models is orders of magnitude less computationally demanding than the dynamical equivalent \citep{Bordoy2014}. In particular they do not require the use of a supercomputer, meaning that they can be employed on demand by users outside of traditional HPC centres for regions of interest. Their relatively low cost also makes it possible to create larger ensembles that would be infeasible using dynamical models. 

The principle behind stochastic downscaling is to use information about the small-scale structural properties of the target field to inform the construction of the downscaled fields. To do stochastic downscaling, we must make some assumption about how the field looks at smaller scales. For instance we might assume that the field looks the same as past data, that it follows a power scaling law, or that it has a known covariance structure.

Assuming similarity with past data leads us to the use of ensemble analogs. The idea is essentially the same as the k-nearest neighbours algorithm: a database of previously observed, high-resolution fields is searched for the fields which most closely resemble the current situation, with similarity based on smooth synoptic variables such as pressure \citep{monache2013probabilistic}. Analog methods have the advantage of being non-parametric, meaning they are easy to apply to variables with arbitrary distributions, and able to capture consistent spatial and multivariate properties \citep{zorita_analog_1999}. However, the analog approach relies on having a large enough data set of potential analogues to be confident of finding a close match, which \cite{van_den_dool_searching_1994} argues is unrealistic. In general, these methods are constrained by the completeness of their data sets; for example, they are unable to predict extremes which lie outside the previously observed range.

An alternative approach is to make use of approximate power-scaling laws which are empirically observed for variables such as precipitation \citep{gupta_statistical_1993}. Power-scaling law assumptions can be used as a basis for various models (including random field models, as in \cite{wikle_spatiotemporal_2001}). One example is multiplicative cascade methods \citep{perica_model_1996}, which are based on repeated use of the Haar wavelet transform \citep{strang_wavelet_1993} to create a field at twice the original resolution by stochastically sampling directional fluctuations. The power law assumption means that, unlike analogs, multiplicative cascade methods can be effective with small datasets. They are also generally successful in simulating realistic distributions of rainfall intensities. However, multiplicative cascade methods do not lend themselves to modelling spatial correlations and tend to introduce artificial discontinuities \citep{gagnon_spatial_2012}. This provides the motivation for methods based on random fields, which directly model the covariance between grid-cells and can better represent spatial structure. 

Gaussian random fields approach downscaling by using an assumed covariance form to infer small-scale structure. As the main topic of this work, they are reviewed in more detail in the following section.

\subsection{Gaussian random fields}

Gaussian random fields (GRFs) are a geostatistical method used in a range of spatial inference problems in fields such as geological sciences \citep{Li2016Profiles} and disease mapping (usually in the form of Gauss Markov random fields - see e.g. \cite{Martinez2013}). They are mathematically equivalent to the machine learning method Gaussian processes \citep{Rasmussen2006Gaussian}, with the former term reserved for models with a spatial interpretation. The term "kriging" is also found in the literature, and is more or less equivalent. GRFs have been applied to various problems in atmospheric science, such as modelling wind \citep{hewer_matern-based_2017, wikle_spatiotemporal_2001}, precipitation \citep{nychka_multiresolution_2015}, and energy production \citep{wytock_large-scale_2013,bei_zhang_spatial-temporal_2016}.

Besides the usual application of interpolation from point observations, Gaussian random fields can be applied in the more general case of observations based on areal averages, sometimes referred to as area-to-point (ATP) or area-to-area (ATA) kriging \citep{Gotway2002, Kyriakidis2004, GE2019102897, HU2020104471}. This broader view has found application to a number of interpolation tasks in disease risk modelling \citep{Kelsall2002}, soil nutrient mapping \citep{Schirrmann2012}, and remote sensing \citep{PARDOIGUZQUIZA200686}. Our approach to spatial conditioning belongs to this family of methods, although with a greater emphasis on stochastic field generation than interpolation.

In the stochastic downscaling paradigm, the most closely related work is that of \cite{allcroft_latent_2003} and \cite{gagnon_spatial_2012} on precipitation downscaling. Both use a Gibbs sampling approach to approximate a conditional GMRF, alternating between a step in which pixels are iteratively resampled based on their neighbours, and a rescaling step which matches the field to observed spatial averages. Our approach differs from these in the explicit use of spatial conditioning, and the use of an adaptive covariance structure.

A challenge of downscaling to convective scales is that the spatial properties of the target field are themselves highly variable \citep{flack_characterisation_2016}. This complicates the task of finding a suitable assumption about the small-scale properties, which is needed for stochastic downscaling. GRFs provide a plausible line of approach on this problem, as a small number of parameters can be used to express a wide range of spatial properties. This approach has not yet been fully utilised in previous work. A time-constant covariance is used for the model in \cite{allcroft_latent_2003}; \cite{gagnon_spatial_2012} also uses a constant covariance structure, but with a variable standard deviation estimated from the large-scale convective available potential energy (CAPE) value.

We here take a different approach, using an adaptive covariance which is estimated from the observed low-resolution field using the maximum likelihood. The intention is to account for the widely varying spatial structures of convective fields, themselves driven by changes in the prevailing synoptic situation. 

\section{Problem setup} \label{setup}

\begin{figure}[h!]
\begin{center}
\includegraphics[width=0.6\columnwidth]{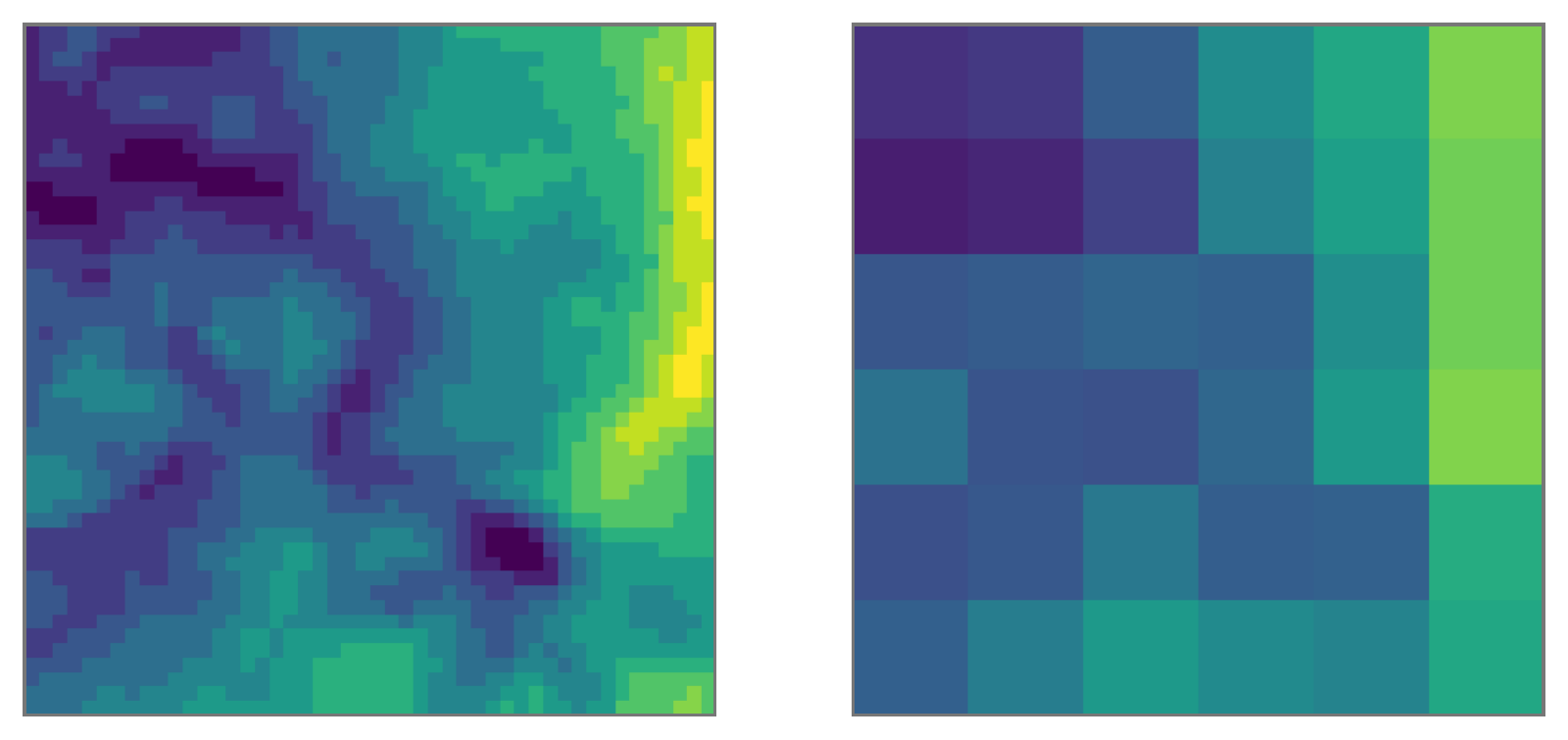}
\end{center}
\caption{Original wet bulb potential temperature field together with synthetic synoptic scale field (right) formed by block averaging with block size $8\times8$.} \label{block}
\end{figure}

In this work, we consider downscaling from synthetically coarse-grained weather fields, produced by block averaging high-resolution fields (Figure \ref{block}). This approach is in line with the bulk mass flux view of convective parameterisation, in which subgrid-scale clouds are treated as an ensemble \citep{arakawa_interaction_1974, gregory_mass_1990}. From the coarse-grained field, our aim is then to generate plausible reconstructions of the original atmospheric conditions. The coarse-graining can be performed with different block sizes to produce fields of different resolutions. 

Our high resolution data is taken from the Met Office operational 2016 Mogreps-UK model, which has a resolution of 2.2-km and so is convection permitting \citep{Golding2016MOGREPS-UKResults}. We then coarse-grain this to two different extents: a synoptic scale of \textasciitilde
20km (equivalent to an $8\times8$ coarsening) and a mesoscale of \textasciitilde
10km (equivalent to a $4\times4$ coarsening).

Our downscaling task is, given a coarse-grained field, to produce samples which match the properties of the original field. In particular, we will compare the following metrics: the mean squared error (MSE), the power-spectral density (PSD), the continuous ranked probability score (CRPS), and a new metric called the neighbourhood Wasserstein score (NWass) (see Section \ref{verification}). Each metric targets a different property of the sample fields, and whether or not it matches the target field. The point-by-point similarity is measured by MSE, the spectral properties such as the overall roughness or smoothness by PSD, the similarity of values over larger areas are compared by NWass, and the distributional properties are measured by CRPS. 

\section{Method} \label{method}

In this paper, we adopt a Gaussian random field approach. Our method has two main components: a linear operation which defines the distribution of the target field conditional on the coarse-grained field, given an estimated covariance of the target field; and an estimation of the covariance of the target field from the coarse-grained field by maximum likelihood. Taken together, these make it possible to construct sample outputs which match the properties of the original field, given a coarse-grained version of the field. 

Before describing our model in detail in Section \ref{mod}, we briefly introduce Gaussian random fields and their parameterisation using kernels.

\subsection{Gaussian random fields} \label{grf}

A Gaussian random field (GRF) is formally defined to be a collection of random variables indexed by one or more spatial variables, such that any finite sub-collection has a multivariate Gaussian distribution \citep{Rasmussen2006Gaussian}. In our context, these variables are our target grid point values, so our geospatial fields are represented by high-dimensional Gaussian distributions with each grid point represented by a different orthogonal dimension. This assumes a Gaussian distribution of values which is  approximately true for the variables we are considering.\footnote{It should be noted that many variables of importance in atmospheric science are not close to Gaussian distributed, notably moist variables such as precipitation and cloud coverage. Applying this method to such variables will likely result in an unrealistic distribution. If the deviation from Gaussianity is relatively minor at small scales, this may still be an acceptable approximation. If the deviation is more significant, it may be possible to apply this method by either normalising the data using a Box Cox transformation \citep{BoxCox} or by viewing the field as a transformation of a latent Gaussian process \citep{Kleiber2012}.}

GRFs have several advantages as a machine learning model, being intrinsically probabilistic and able to flexibly incorporate expert knowledge through the form of the mean and covariance  \citep{Rasmussen2006Gaussian}. In addition, GRFs have convenient mathematical properties which make it possible to evaluate many quantities of interest analytically. Conditioning the distribution on data and drawing samples are important examples of this, both of which reduce to simple matrix algebra in the Gaussian case. We describe how our model makes use of these properties in Section \ref{condsamp}. 

A disadvantage of GRFs is that they can be computationally demanding for large datasets, scaling cubically with the number of sample points. Computational efficiency is an important consideration for statistical downscaling models, but this limitation does not excessively impact us in the present use-case.

\begin{figure}[ht!]
\begin{center}
\includegraphics[scale=1]{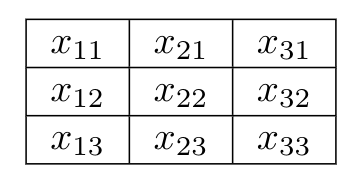}
\end{center}
\caption{Grid on which we may define a Gaussian random field.} \label{grid}
\end{figure}

As an example of a GRF, consider the finite case of a $3 \times 3$ spatial grid $X$ shown in Figure \ref{grid}. Then a Gaussian field $G$ defined over $X$ is a 9-dimensional Gaussian distribution with mean $\mu$ and covariance $C$ given by

\begin{equation*}
  \mu = \begin{bmatrix}
        \mu_{11} \\
        \mu_{12} \\
        \vdots \\
        \mu_{33}
		\end{bmatrix} \qquad
  C = \begin{bmatrix}
        \mathrm{cov}(x_{11},x_{11}) & \mathrm{cov}(x_{12},x_{11}) & \cdots & \mathrm{cov}(x_{33},x_{11})\\
        \mathrm{cov}(x_{11},x_{12}) & \mathrm{cov}(x_{12},x_{12}) & \cdots & \mathrm{cov}(x_{33},x_{12}) \\
        \vdots & & \ddots \\
        \mathrm{cov}(x_{11},x_{33}) & \mathrm{cov}(x_{12},x_{33}) & \cdots & \mathrm{cov}(x_{33},x_{33})
		\end{bmatrix}.
\end{equation*}
\

\begin{figure}[ht!]
\begin{center}
\includegraphics[width=\columnwidth]{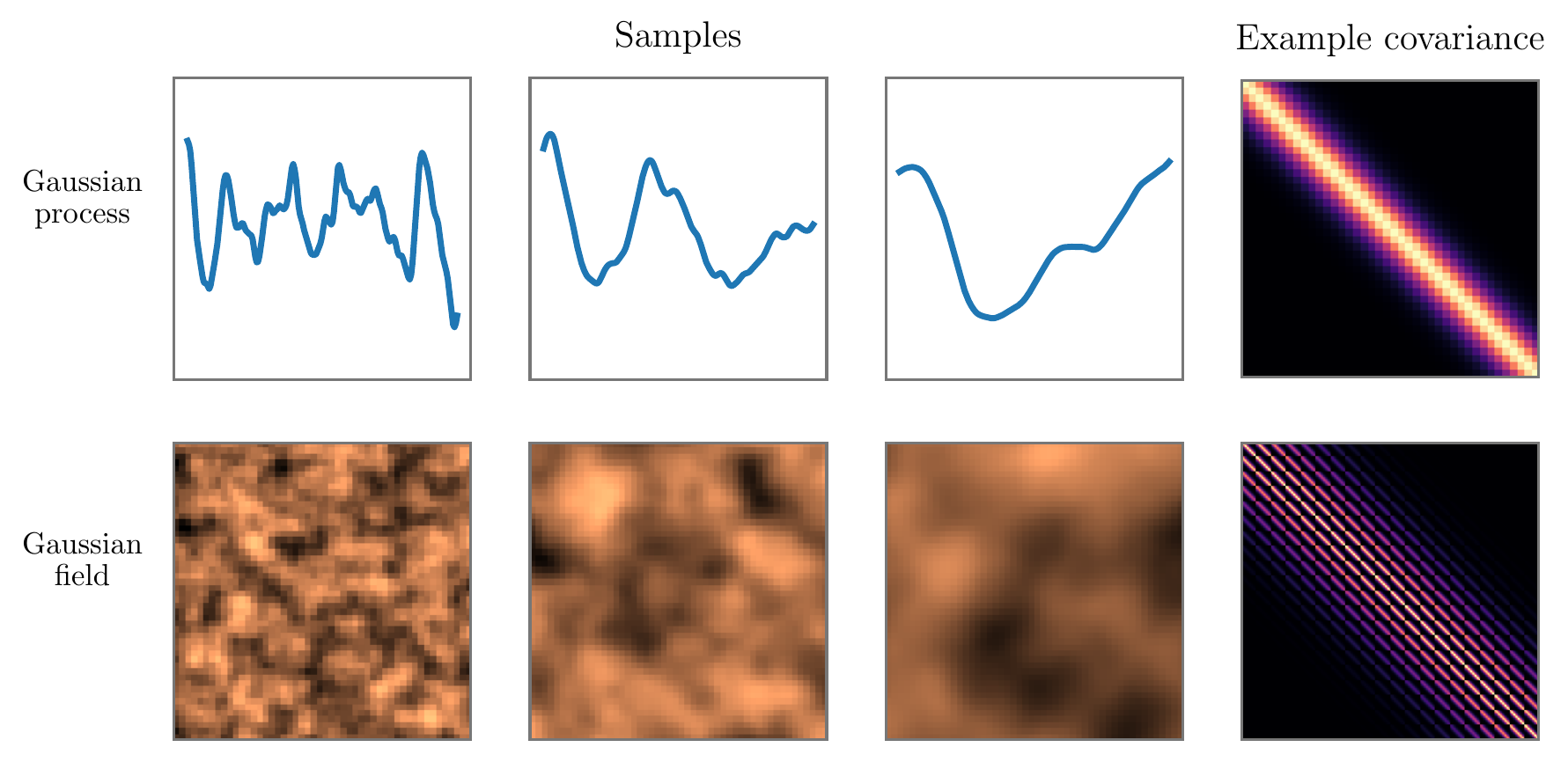}
\end{center}
\caption{Examples of 1D Gaussian processes and 2D Gaussian random fields. The right-hand column shows the structure of the covariance matrix for each dimension. Other columns show samples drawn using different kernel lengthscale parameters.} \label{grf_examples}
\end{figure}

The mean can be any real-valued 9-dimensional vector. Theoretically, the covariance can be any valid $9 \times 9$ covariance matrix; that is, it must only satisfy the conditions of being symmetric and positive semi-definite. However, in standard practice Gaussian random fields are not arbitrary Gaussian distributions; they have spatial structure. Figure \ref{grf_examples} shows how this spatial structure can be encoded in the covariance matrix. For a single dimension, the structure manifests as a concentration of positive covariance near the diagonal, indicating that nearby points should be closely related. The structure for two dimensions is harder to interpret, due to the two dimensions being flattened into a single index, but the interpretation is similar: any point is most closely related to its close neighbours in two dimensions.

The covariance is generally taken to have one of a number of standard forms known as kernel functions (or covariance functions, or covariance kernels), which are known to form valid covariance matrices. Kernel functions are discussed in detail in \cite{Rasmussen2006Gaussian}. Using kernels to describe covariance matrices dramatically simplifies covariance estimation. Instead of having to estimate $N^4$ parameters for a 2-dimensional $N \times N$ space, it is only necessary to learn the kernel parameters, of which there are usually just one or two. We make use of this convenient estimation property in Section \ref{fit}. The effect of kernel parameters is illustrated in Figure \ref{grf_examples}, which shows the effect of using different lengthscale parameters.

Kernels are often selected to encode properties which are believed to hold for the system under consideration. The standard assumptions are:

\begin{enumerate}
    \item \textbf{locality}: covariance decreases with distance, that is nearby points are more closely related than distant points.
    \item \textbf{isotropy}: the covariance of two points only depends on the distance between them, not on the direction.
    \item \textbf{spatial stationarity}: the covariance of two points depends only on the relative distance between them, not on the absolute location of either point.
\end{enumerate}

\noindent These assumptions apply to all spatial models. If we are considering spatio-temporal models, we can add a fourth assumption:

\begin{enumerate}
\setcounter{enumi}{3}
    \item \textbf{temporal stationarity}: the covariance between two points does not change with time.
\end{enumerate}

While these assumptions considerably simplify the analysis, their validity in atmospheric science is far from guaranteed. For instance, global teleconnections would break the locality assumption, continental versus marine air masses would break the isotropy assumption, orographic forcing would break the spatial stationarity assumption, and seasonality would violate the temporal stationarity assumption. Fronts would be another example, as they can introduce sharp discontinuities which would be excessively smoothed over by standard kernels. However, these assumptions can often be lessened or removed by using more flexible models for the covariance, at the cost of increasing complexity. 

\subsection{An adaptive GRF model for downscaling} \label{mod}

The challenge of downscaling to convective scales is that the covariance structure is highly variable. Motivated by this problem, in this paper we use a time-varying covariance function which can adapt to these changing conditions. 

By fitting to each time step separately, we reject assumption four, that is we allow relationships between points to change with time. In contrast, since we only target a small area in the present work, the effect of different regimes should be minimal within a given timestep. The structure of the downscaling problem itself works to justify the assumption of locality, which need only be assumed conditional on the coarse-grained field. 

\begin{figure}[ht!]
\centering
\begin{tikzpicture}
  [node distance=.5cm,
  start chain=going below,
  scale=0.7, every node/.style={scale=0.7}]
     \node[punktchain, join] (lr) {Low res input};
     \node[punktchain, join] (cov)      {Estimate covariance};
     \node[punktchain, join] (post)      {Condition on input};
     \node[punktchain, join] (hr)      {Draw high res samples};
     
  \draw[|-,-|,-, thick,] (lr.east) -| (3,0);
  \draw[|-,-|,-, thick,] (3,0) |- (3,-3.6);
  \draw[|-,-|,->, thick,] (3,-3.6) -- (post.east);
\end{tikzpicture}
\caption{General framework for adaptive probabilistic downscaling. Information from the low resolution input field is used to estimate the covariance; the estimated covariance and input are then combined to obtain a conditional distribution. This is carried out independently for each time step, eliminating the assumption of temporal stationarity.} \label{framework}
\end{figure}
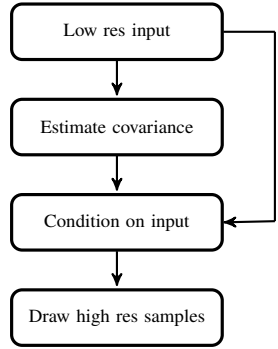

The setup of our model is as follows. To capture the dependence of the covariance structure on weather conditions, a separate covariance model will be required for each timestep, estimated from low-resolution data. We can then condition on the low-resolution field to give our final model. The general framework is illustrated in Figure \ref{framework}. 

This approach gives us the advantages of GRFs discussed previously but additionally should allow us to capture time varying properties of changing weather regimes. Note as well the absence of a training stage - the length scale, variance, and covariance of the high resolution fields are estimated from that of the low resolution field, allowing us to generate high resolution fields.

\subsubsection{Conditioning and sampling} \label{condsamp}

Conditioning a Gaussian random field on a subset of its values, or on a linear combination, reduces to closed-form matrix algebra by means of the relations:

\begin{align}
\boldsymbol{\mu}_{t|o} &= \boldsymbol{\mu}_{t} + C_{t,o}^T \cdot C_{o}^{-1} \cdot (\mathrm{obs} - \boldsymbol{\mu}_{o}) \\
C_{t|o} &= C_t - (C_{t,o}^T \cdot C_{o}^{-1} \cdot C_{t,o})
\label{cond_eqn}
\end{align}

\noindent where a subscript $\cdot_o$ denotes observed variables (variables on which we condition), and a subscript $\cdot_t$ denotes target variables (for which we obtain a conditional distribution). A proof of these relations is given in Appendix \ref{cond_proof}. Here, $C_{t|o}$ is the covariance of the target variables conditional on the observations, $C_t$ is the unconditioned covariance of the target field, $C_{o}$ is the covariance of the observed variables, and $C_{t,o}$ is the joint covariance of the target and observed fields. Likewise, $\boldsymbol{\mu}_{t|o}$ is the conditional mean of the target field, $\boldsymbol{\mu}_{t}$ is the unconditioned mean, $\boldsymbol{\mu}_{o}$ is the expected value of the observed variables, and $\mathrm{obs}$ is the true observation.

We can apply this to our downscaling problem by noting that our coarse-graining operation can be represented by a matrix multiplication. For a coarse-graining matrix $A$, we have the linear constraint $A\mathbf{x = \overline{x}}$ where $\mathbf{\overline{x}}$ is the coarse-grained field. Following \cite{Rue2005GaussianApplications}, the joint distribution of $\mathbf{x}$ and $A\mathbf{x}$ is given by

\begin{equation}
\mathbb{E}\begin{pmatrix}
\mathbf{x} \\
A \mathbf{x}
\end{pmatrix} =
\begin{pmatrix}
\boldsymbol{\mu}_t \\
\boldsymbol{\mu}_o
\end{pmatrix} = 
\begin{pmatrix}
\boldsymbol{\mu}_t \\
A \boldsymbol{\mu}_t
\end{pmatrix} \qquad
\mathrm{Cov}\begin{pmatrix}
\mathbf{x} \\
A \mathbf{x}
\end{pmatrix} =
\begin{pmatrix}
C_t & C_{o,t} \\
C_{t,o} & C_o
\end{pmatrix} =
\begin{pmatrix}
C_t & C_tA^T \\
A C_t & A C_t A^T
\end{pmatrix}
\end{equation}

\noindent Using the relations of equation \ref{cond_eqn}, the distribution $p(\mathbf{x} | A\mathbf{x} = \mathbf{\overline{x}})$ can then be found by computing

\begin{align}
\boldsymbol{\mu}_{t|o} &= \boldsymbol{\mu}_t - C_tA^T(AC_tA^T)^{-1}(\mathbf{\overline{x}} - A\boldsymbol{\mu}_t ) \\
C_{t|o} &= C_t - C_tA^T(AC_tA^T)^{-1}AC_t.
\label{eq:cond}
\end{align}

\noindent Proof of these identities is given in Appendix \ref{cond_proof}.. Now, $\boldsymbol{\mu}_{t|o}$ and $C_{t|o}$ are the mean and covariance of the conditional GRF given by conditioning on a low-resolution "observation". This forms the basis for our probabilistic downscaling model. 
This conditioning relies on having an estimate of the covariance of the high resolution field, given by $C_t$. Estimating $C_t$ is the subject of the following Section. 

Having obtained our conditional distribution, we can draw samples using the standard method for sampling a Gaussian distribution: if $\mathbf{w}$ is a vector containing white noise (Gaussian with zero mean and unit variance) then 

\begin{equation}
\mathbf{s} = \boldsymbol{\mu}_{t|o} + {C_{t|o}}^{\frac{1}{2}} \mathbf{w}
\end{equation}

\noindent is a sample from our conditional distribution, where for a matrix $M$ the symbol $M^{\frac{1}{2}}$ denotes a matrix with the property that 
$
M = M^{\frac{1}{2}}M^{{\frac{1}{2}}T}.
$

\paragraph{Example of spatial conditioning}

Since our approach to conditioning is different to the standard approach used in the Gaussian process and GRF literature we will demonstrate using a one-dimensional example. 

\begin{figure}[ht!]
\begin{center}
\includegraphics[width=0.5\columnwidth]{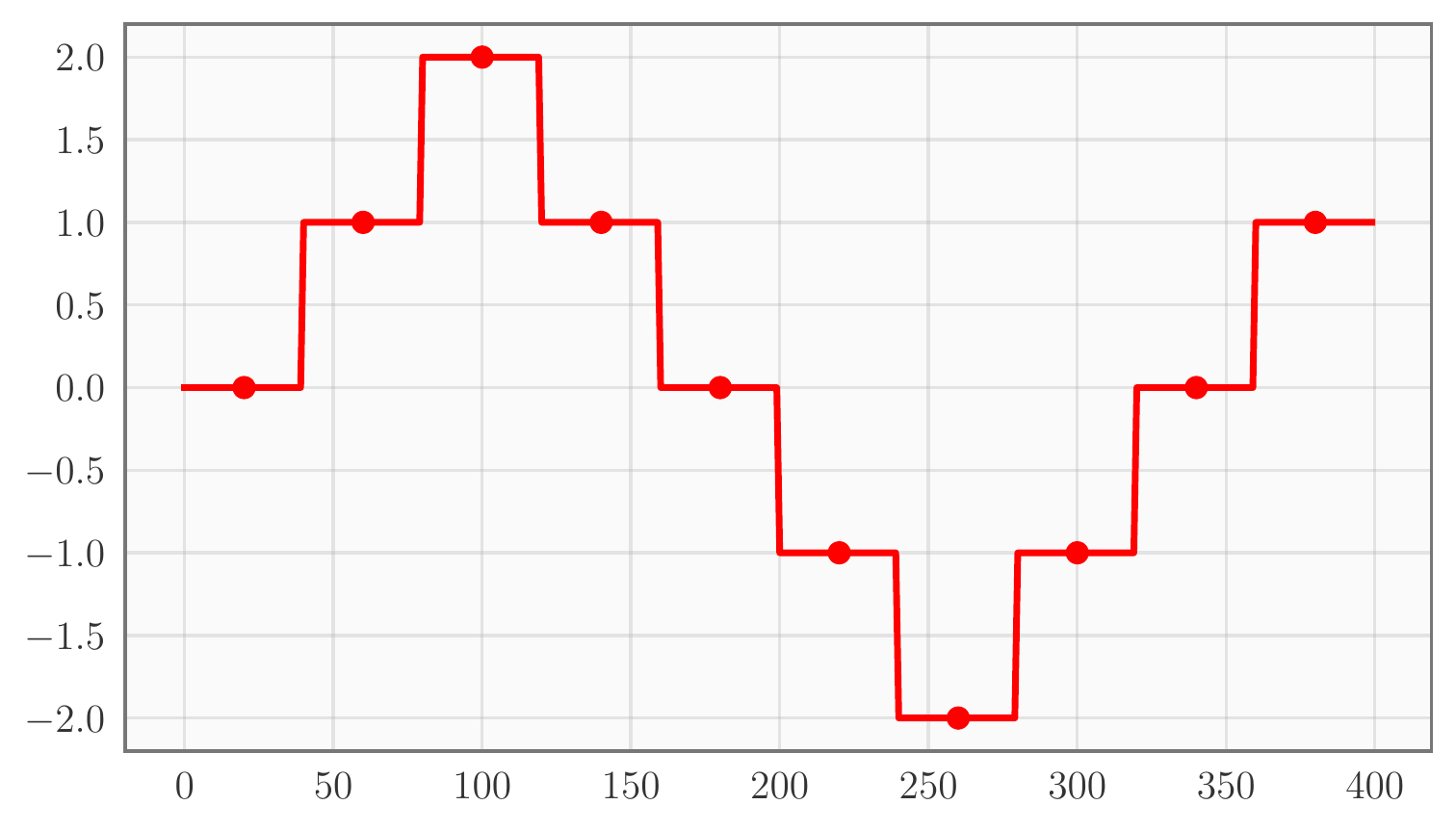}
\end{center}
\caption{Observed values. Dots indicate an interpretation as point observations, horizontal lines indicate an interpretation as spatial averages.} \label{1d_obs}
\end{figure}

Consider a sequence of observations which we will use to condition a Gaussian process model (Figure \ref{1d_obs}). Depending on what we know about the system we are trying to model, we might interpret this sequence of observations as point values or as spatial averages. 

\begin{figure}[ht!]
\begin{center}
\includegraphics[width=1\columnwidth]{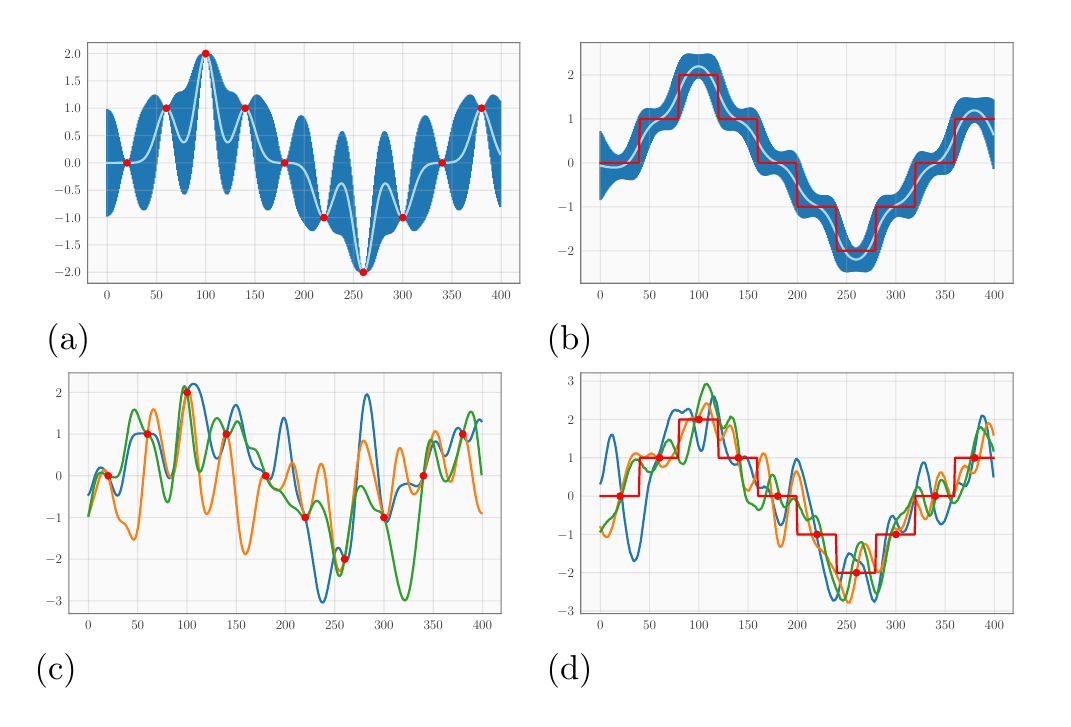}
\end{center}
\caption{The distribution obtained by conditioning on point values (a), and spatial averages (b). Sample fields are shown beneath the corresponding distributions in (c) and (d). The distributional means are shown in pale blue, and the marginal variance in darker blue.}
\label{1d_cond}
\end{figure}

The effects of this choice of interpretation are shown in Figure \ref{1d_cond}. The prior mean, variance, and covariance are the same in each case (specifically we use zero mean, unit variance, and squared exponential covariance with lengthscale ten). However, the two conditional distributions have quite different properties. Conditioning on point values leads to a ``pinched'' marginal variance with much higher variance at a distance from the observations. By contrast, conditioning on spatial averages leads to a more uniform level of variance. 

\begin{figure}[ht!]
\begin{center}
\includegraphics[width=0.5\columnwidth]{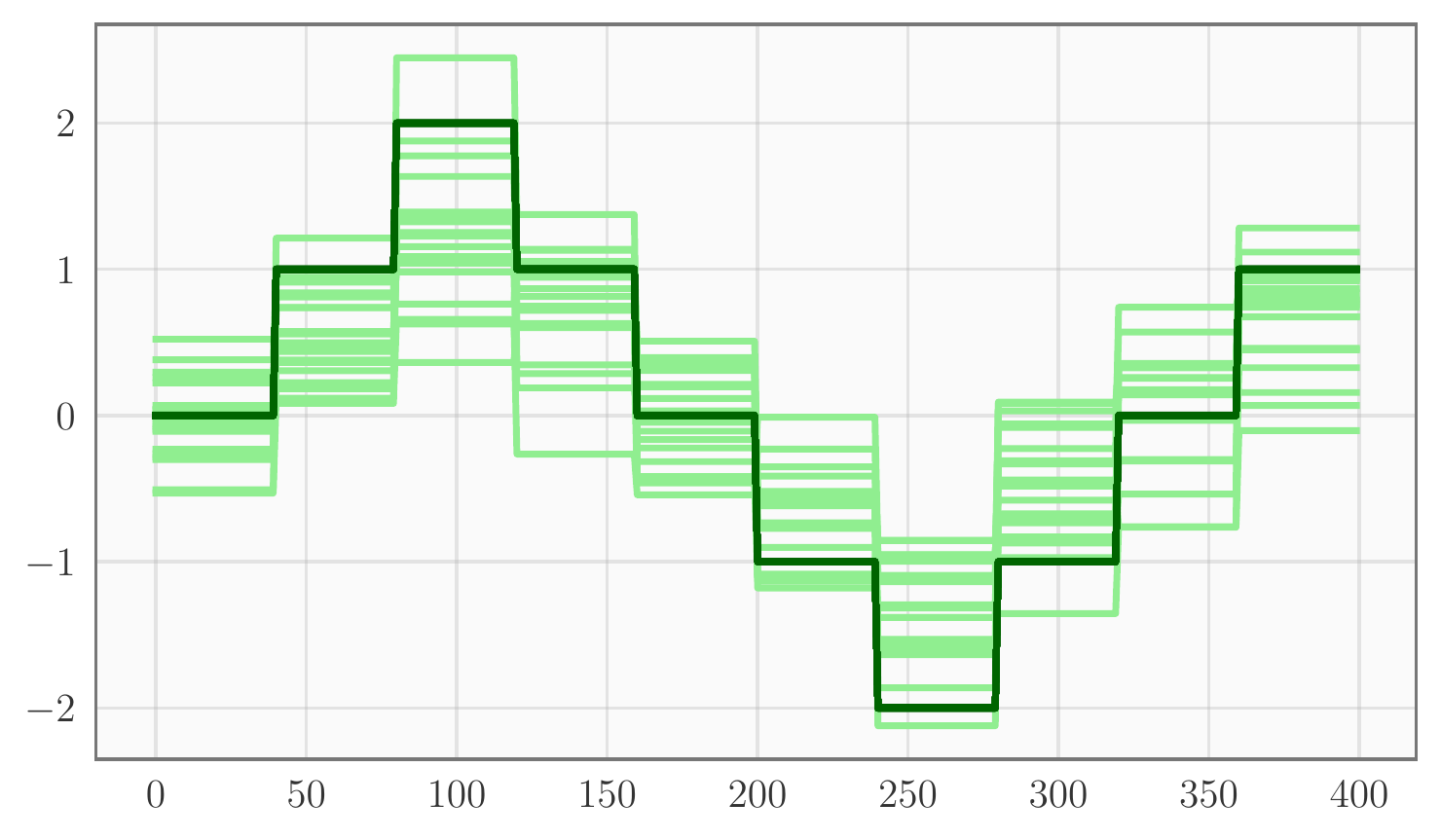}
\end{center}
\caption{Spatial averages of samples from the distribution conditioned on spatial averages (dark green) and point observations (light green).} \label{1d_emp_sp}
\end{figure}

While the sample fields in Figure \ref{1d_cond}(c) are constrained to pass through the observed points, the constraint on the sample fields shown in Figure \ref{1d_cond}(d) is less easy to see. However, Figure \ref{1d_emp_sp} makes this constraint easy to see. It shows that the spatial averages of the sample fields conditioned on spatial averages are all in agreement with the original coarse-grained observations, while those conditioned on point values show substantial variability. This property is enforced directly by conditioning on these spatial averages, as described in equation \ref{eq:cond}. Instead of conditioning on a specific point value, we have effectively conditioned on a low resolution version of the field - just as required for our downscaling task.

\begin{figure}[ht!]
\begin{center}
\includegraphics[width=0.8\columnwidth]{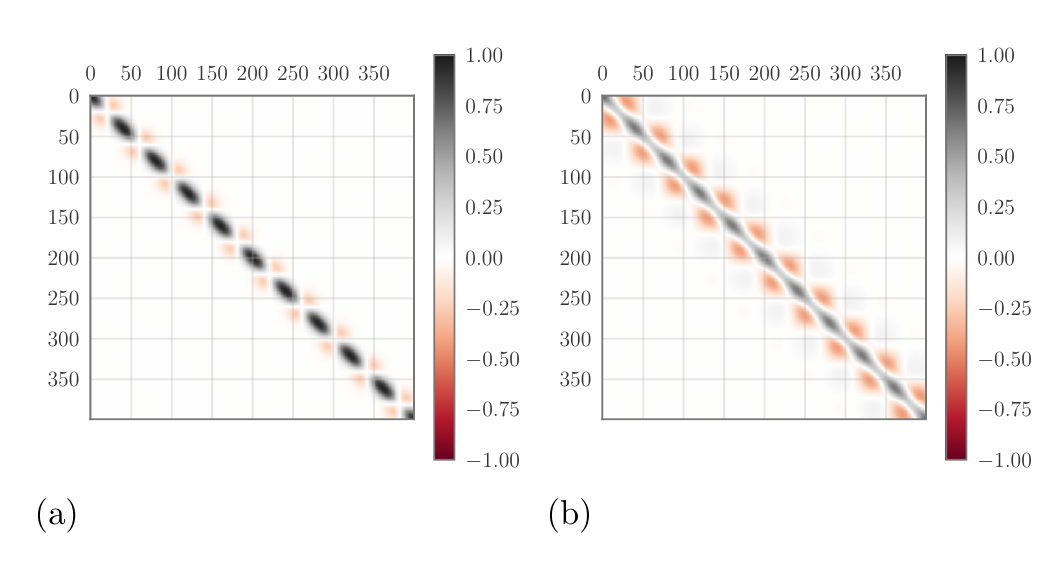}
\end{center}
\caption{Posterior covariance matrices after conditioning on point observations (a) and spatial averages (b).}  \label{1d_post_c}
\end{figure}

As a further illustration, notice how the properties of the two distributions are encoded in the posterior covariance matrices shown in Figure \ref{1d_post_c}, particularly in the distribution of their negative values. Conditioning on a point observation introduces a localised positive/negative ``dipole'' forcing smoothness of the samples through the observation. On the other hand, conditioning on a spatial average induces a wider-scale negative correlation between the points in the area being averaged. This is in contrast to the prior covariance, which consists of only positive values which peak along the diagonal (not shown).

All the essential properties of this example hold true in two or more dimensions; we have chosen here to illustrate the idea with one dimension for ease of interpretation. The rest of the paper is focused primarily on the two-dimensional case.


\subsubsection{Covariance estimation} \label{fit}

In order to perform the conditioning step of equation \ref{eq:cond} we need to have an estimate of the high resolution covariance $C$. As discussed above, since this estimated covariance is what controls the behaviour of our model at small scales, it needs to be estimated separately for each timestep. Fortunately, by making use of a parameterised covariance kernel we can reduce this to a low dimensional optimisation problem.

We have chosen to consider the Matern covariance due to its flexibility in modelling smoothness properties \citep{stein_interpolation_1999}. We treat the shape parameter $\nu$ as a hyperparameter, leaving us to fit two parameters for each input: the lengthscale $\ell$ and the variance $\sigma^2$. 

The lengthscale $\ell$ we estimate using maximum likelihood estimation, where the likelihood of a parameter is defined to be the joint density of the observed data as a function of the parameter only, with the data held fixed. The density for a GRF is just the standard Gaussian density, where we take the mean to be zero and the covariance $K(\theta)$ to be determined by its parameters $\theta$:

\begin{equation}
p(\mathbf{x}_i | \theta) = \frac{1}{\sqrt[]{2\pi |K|}}\ \mathrm{exp}(-\tfrac{1}{2}\mathbf{x}_i^T K^{-1} \mathbf{x}_i)
\end{equation}

\noindent thus the log likelihood is given by 

\begin{equation}
\mathrm{log}\ p(\mathbf{x} | \theta) = -\frac{1}{2}\bigg(n\ \mathrm{log}(2\pi) + n\ \mathrm{log}|K| + \sum_{i=1}^n \mathbf{x}_i^T K^{-1} \mathbf{x}_i\bigg).
\end{equation}

\noindent By maximising the log likelihood using standard optimisation methods, we can find the kernel parameter values which best fit the observed data. Thus we approximate the true target covariance $C_t$ of equation \ref{eq:cond} by a parameterised version $K(\theta)$.

In the conditional downscaling task we do not observe samples of the target field directly, but only after convolution with a filter matrix $A$. Therefore, the relevant covariance matrix is not the one computed directly from the kernel, $K(\theta)$, but the filtered covariance $AKA^T$. The log likelihood is then given by

\begin{equation}
\mathrm{log}\ p(\mathbf{x} | \theta) = -\frac{1}{2}\bigg(n\ \mathrm{log}(2\pi) + n\ \mathrm{log}|K||A|^2 + \sum_{i=1}^n \mathbf{x}_i^T (AKA^T)^{-1} \mathbf{x}_i\bigg)
\end{equation}

\noindent which gives an analogous optimisation.

The optimisation of the variance can be separated from the optimisation of the other parameters. This is possible because the dependence of $K$ on the variance $\sigma$ is just $K_{\sigma} = \sigma K_1$. We then have $K_{\sigma}^{-1} = \frac{1}{\sigma} K_1^{-1}$ and $\frac{\partial K_{\sigma}}{\partial \theta_j} = \sigma \frac{\partial K_1}{\partial \theta_j}$. Any dependence on $\sigma$ will therefore drop out of the gradient equation, so that the extremal values of any other parameters are independent of the variance. The variance parameter can then be optimised in a separate step.

Taken together, the methods described in this Section allow us to estimate the full covariance matrix for a high resolution field given a coarse-grained version of that field. We can then apply the conditioning and sampling steps of Section \ref{condsamp} to generate plausible reconstructions.

\section{Verification} \label{verification}

The question of how to best evaluate probabilistic downscaling models is a challenging one, especially when the target structure is chaotic. This reflects a known difficulty in verifying high resolution physics-based models; although there are sound reasons to believe that increased resolution improves the realism of the model, this improvement is often not reflected by standard verification scores \citep{Sansom2015AdvancesVerification}. In particular, point-based verification metrics such as mean-squared error (MSE) have the undesirable property that insignificant displacement of features is heavily penalised due to the double-penalty problem: the displacement is treated as a missing feature, and a separate anomalous feature. As has been observed in the context of machine learning, such scores tend to reward blurry predictions and penalise specificity \citep{Mathieu2015DeepError}.

In order to address these difficulties, we consider two additional deterministic verification metrics besides the standard MSE. These are the spectral and neighbourhood Wasserstein scores, which are both spatial verification metrics applied to samples drawn from the probabilistic model. We also consider the distribution properties of stochastic methods by means of the continuous ranked probability score (CRPS).

\subsection{Spectral score}

The purpose of the spectral score is to evaluate the quality of the generated fields in frequency space as opposed to physical space. The generated and target fields are first transformed into Fourier space using the two-dimensional Fourier transform. The one-dimensional power spectral density (PSD) is then given by the squared amplitude component integrated over circles of given radius. For a given radius $\varphi$ we have:

\begin{equation}
\mathrm{PSD}_\varphi(\mathbf{x}) = \mathlarger{\int}\displaylimits_{\sqrt{\omega^{2} + \sigma^{2}} = \varphi} \hat{\mathbf{x}}^2 (\omega, \sigma)\ \mathrm{d}\ \mathrm{tan}^{-1}\bigg(\frac{\sigma}{\omega}\bigg).
\end{equation}

\noindent where the integral is taken over the angle $\theta = \mathrm{tan}^{-1}(\frac{\sigma}{\omega})$ and $\mathbf{\hat{x}}$ denotes the two-dimensional Fourier transform of $\mathbf{x}$. We will use the notation $\mathrm{PSD}(\mathbf{x})$ to denote the power spectral density as a function of $\varphi$, so that $\mathrm{PSD}(\mathbf{x})(\varphi) = \mathrm{PSD}_\varphi(\mathbf{x})$.

The generated and target fields can then be compared using any of the standard histogram distance measures. As in \cite{Martinez2016AData}, we here use the 1-Wasserstein distance between their PSDs. The $p$-Wasserstein distance is defined to be

\begin{equation}
W_p(\mu, \nu) = \bigg(\inf\limits_{\pi \in \Pi(\mu, \nu)} \displaystyle\int_{M \times M} d(x, y)^p \ \mathrm{d}\pi(x, y)\bigg)^{\frac{1}{p}}
\label{eq:wass}
\end{equation}

\noindent where $\mu$ and $\nu$ are probability measures on $M$, and $\Pi(\mu, \nu)$ is the collection of all measures on $M \times M$ with marginals $\mu$ and $\nu$ \citep{villani2008optimal}. Where $p=1$, the Wasserstein distance is the same thing as the earth mover's distance, or optimal transport, which describes the cost of transforming one distribution into another \citep{arjovsky_wasserstein_2017}. This is particularly appropriate for comparing power spectra, as the cost of small frequency displacements is correspondingly small. 
The spectral score is then given by

\begin{equation}
W_1(\mathrm{PSD}(\mathbf{x}), \mathrm{PSD}(\mathbf{x^*}))
\end{equation}

\noindent where $\mathbf{x}$ and $\mathbf{x^*}$ are the target and generated fields respectively, and we have abused notation to write $\mathrm{PSD}(\mathbf{x})$ to denote the image of that function.

\subsection{Neighbourhood Wasserstein score} \label{nwass}

Point-based verification methods have the undesirable property that slight displacement of features is heavily penalized due to the double-penalty problem: the displacement is treated as a missing feature, and a separate anomalous feature. In contrast, spatial verification methods are designed to avoid, or at least characterize, this effect. The main sub-categories of spatial verification include scale-separation approaches (e.g. \cite{Casati2004}), feature-based methods (e.g. \cite{Davis2006}), and neighbourhood methods. Neighbourhood methods are especially well-suited to our purposes. Unlike feature-based methods they can be easily applied to continuous fields (e.g. \cite{Rezacova2007AStorms}). Scale-separation methods are designed to assess the scale-dependence of errors, but do not in themselves address the double penalty problem as do neighbourhood methods \citep{YU2020105117}.

The assumption underlying neighbourhood verification methods is that a slightly displaced forecast may still be more useful than one which fails to capture the relevant features \citep{Ebert2008FuzzyFramework}. Instead of comparing fields point by point, these methods consider a neighbourhood surrounding each gridcell and some measure of similarity of the real and predicted neighbourhoods. The canonical example is the fractions skill score \citep{Roberts2008ScaleEvents} which compares the number of grid cells exceeding a given threshold within a neighbourhood; other variants are reviewed in \cite{Ebert2008FuzzyFramework}. 

When considering a continuous valued field for which there is no obvious threshold of special interest, it is more natural to compare the full distribution of values within a neighbourhood than an exceedance count. One possible score of this type was proposed by \cite{Rezacova2007AStorms}, who use the mean squared distance of vectors containing the ordered values from the two fields. However, there are a number of possible measures of the distance between distributions; several possibilities are discussed in \cite{Bellemare2018TheGradients}, \cite{Ramdas2015OnTests} and \cite{feydy2019interpolating}. We here propose a spatial score which uses the 1-Wasserstein distance, as described in the previous Section (equation \ref{eq:wass}), rewarding fields which have similar distributions over small neighbourhoods. We will call this the neighbourhood Wasserstein score.

\begin{figure}[ht!]
\begin{center}
\includegraphics[scale=0.4]{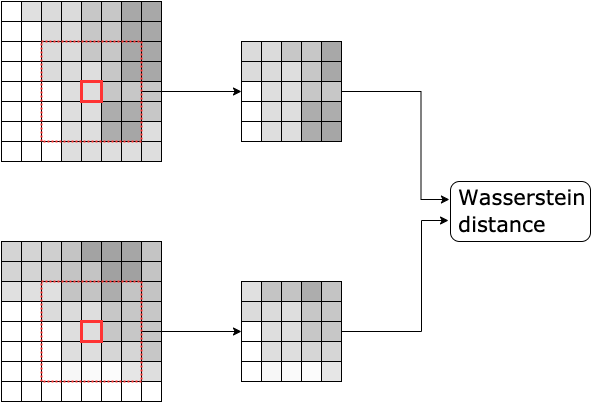}
\end{center}
\caption{Schematic showing calculation of the neighbourhood Wasserstein score.} \label{n_wass}
\end{figure}

Note that we compute the 1-Wasserstein distance between the histogram of values within the neighbourhoods. Thus, the calculation is invariant to the spatial distribution of values within a particular neighbourhood. Computationally, calculating the 1-Wasserstein distance is equivalent to sorting the flattened list of values within the neighbourhoods to be compared, and summing the unsigned differences of the sorted lists. A proof of this equivalence is given in Appendix \ref{wass_appx}.

To compute this score, it is first necessary to choose a neighbourhood size, which sets the side length of a square centered at the target point. For each point, a neighbourhood is extracted from the two fields being compared. The score for that point is the Wasserstein distance between the observed and target neighbourhoods. The overall score is then the mean of the scores for each point:

$$\mathrm{NWass}_k(X, Y) = \frac{1}{N_k}\sum_{n=1}^{N_k}W_{1}(X_n^k, Y_n^k)$$.

\noindent where there are $N_k$ neighbourhoods of size $k$, and $X_n^k$ is the $n$th neighbourhood of size $k$ in $X$. A schematic illustration is shown in Figure \ref{n_wass}.

For any neighbourhood score, it is important to ensure that the choice of neighbourhood size is appropriate to the application. In our experiment, we chose a neighbourhood size of 4x4 grid cells, or approximately 10km. This is slightly below the 15km used in \cite{Golding2016MOGREPS-UKResults}, which justifies that our choice of neighbourhood size is not unreasonably large.

\subsection{Continuous ranked probability score} \label{crps}

While the spectral and neighbourhood Wasserstein scores are able to compare the spatial structure of the forecast and target fields, they are still fundamentally deterministic scores and do not take the full distribution of stochastic models into account. We therefore need to complement these scores with a probabilistic metric to get a full picture of the performance.

For this purpose, we have selected the continuous ranked probability score (CRPS). For a probabilistic forecast, the CRPS is defined as the integrated squared difference between the forecast CDF $F$ and the observed CDF:

$$\mathrm{CRPS}(F, x) \int_{-\infty}^{\infty}(F(y) - \mathbf{1}(y \geq x))^2 \ \mathrm{d}y$$

\noindent where $\mathbf{1}(y \geq x)$ is an indicator function for the observed value $x$ exceeding the threshold $y$ \citep{gneiting2007probabilistic}. 

\section{Application to wet bulb potential temperature} \label{app}

\begin{figure}[ht!]
\begin{center}
\includegraphics[width=0.7\columnwidth]{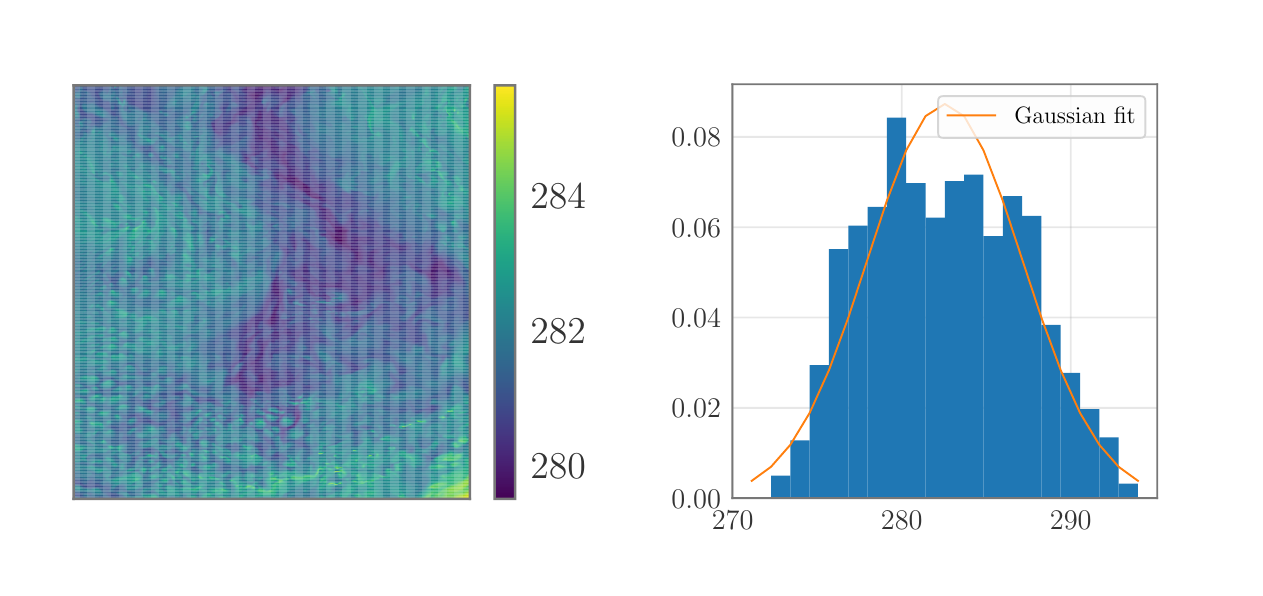}
\end{center}
\caption{Left is an example WBPT field. Right shows histogram of WBPT values for the period under consideration with best fit Gaussian PDF shown in orange.} \label{wbpt_field}
\end{figure}

In this Section, we test the performance of the conditional downscaling model on fields from the Mogreps-UK model produced by the Met Office \citep{Golding2016MOGREPS-UKResults, Hagelin2017TheMOGREPS-UK}. We have chosen to work with wet bulb potential temperature at 850 hPa for two reasons. First, it has substantial small-scale chaotic structure, making it a meaningful target for probabilistic downscaling. Indeed, Table \ref{cloud_corr} shows that WBPT is correlated with low type cloud coverage, particularly at smaller scales. However, unlike cloud coverage its values are approximately Gaussian distributed (Figure \ref{wbpt_field}), making it a more suitable target for our model.


\begin{table}
\begin{center}
 \begin{tabular}{| c | c | c | c |} 
 \hline
  & full & low pass & high pass\\ [0.5ex] 
 \hline
 corr(wbpt,cloud) & 0.32 & 0.27 & 0.38 \\ 
 \hline
\end{tabular}
\end{center}
\caption{Correlation coefficients for wet bulb potential temperature at 850 hPa and low type cloud coverage. Low pass filter is a Gaussian filter with a sigma coefficient of 128, which is subtracted from the original field to give the high pass filter.}
\label{cloud_corr}
\end{table}

Our dataset consists of wet bulb potential temperature data at 850 hPa from the operational 2016 Mogreps-UK model. Dates used are summarised in Table \ref{dataset}. All data is taken from the unperturbed zeroth ensemble member at the three hour timestep of the 15:00 model run. 

To test our model we have chosen to focus on areas of low orography, as this is where stochastic variability is more likely to dominate over orographically-driven variability at small scales. We consider three locations for our dataset, one in South-East England (51.3 - 52.2 latitude, -1.0 - 0.6 longitude), one in the Midlands (51.5 - 52.4 latitude, -2.0 - -0.4 longitude), and one in Ireland (52.2 - 53.2 latitude, -8.4 - -6.8 longitude).

\begin{table}
\begin{center}
 \begin{tabular}{| c | c | c | c | c | c | c |} 
 \hline
 & years & months & days & hours & lead times & members \\ [0.5ex] 
 \hline
 dev & 2016 & 1-12 & 1,11,21 & 15 & 3 & 0 \\ 
 \hline
 dev2 & 2016 & 1-12 & [1-29] $\setminus$ \{5,15,25\} & 15 & 3 & 0 \\ 
 \hline
 eval & 2016 & 1-12 & 5,15,25 & 15 & 3 & 0 \\ 
 \hline
\end{tabular}
\end{center}
\caption{Data used in development and evaluation sets. The dev set is used to fit the shape hyperparameter for the Matern kernel, and the dev2 set is used to for training and hyperparameter tuning benchmark models.}
\label{dataset}
\end{table}

\subsection{Benchmarks}

In order to gauge the effectiveness of using an adaptive covariance, we contrast two versions of our GRF model: one baseline version using a stationary covariance estimated from the full dataset, and one using a seperate covariance estimate for each individual field. We refer to these models as GRF-S (stationary) and GRF-T (time-adaptive). We also consider a third version, identical to GRF-T except that the input variables are treated as point values instead of spatial averages (GRF-T pt); this is to test our assertion that spatial conditioning is more appropriate for our downscaling problem.

We compare our model against five benchmarks. Two are deterministic, and utilise only the low resolution input. These are the synthetic low resolution field upsampled to match the resolution of the target field (lres); and the bicubic interpolation of the low resolution field (bicub). The third is a data-driven deterministic model, namely an ElasticNet regression. Finally, we compare our approach to two stochastic data-driven models based on the cascade decomposition approach.

\subsubsection{ElasticNet model}

For our deterministic data-driven benchmark, we have chosen to use a linear regression mapping the low-resolution covariates to the high-resolution target variables. Specifically, we consider a mapping $Y = f(X)$ from the full low-resolution input $X$ to the full high-resolution output $Y$. Since we are dealing with a large number of dimensions and a relatively modest sized training set, it is important to assess the need for model regularisation. There are three widely adopted methods for regularising linear models: the Lasso model uses an L1 penalty to encourage sparsity in the model weights; Ridge regression uses an L2 penalty to encourage smaller weights; and ElasticNet encompasses both cases by incorporating both an L1 and an L2 penalty \citep{friedman2001elements}. We have chosen the ElasticNet model as it is the most general.

We tuned the model hyperparameters using 5-fold cross validation with a maximum of 10,000 iterations. Following the convention used in the SciKit-Learn library \citep{scikit-learn}, we define L1-ratio to be the relative weighting of the L1 to L2 penalty, and alpha to be a scalar factor multiplying both penalties. Overall, the combination of alpha=0.01 and L1-ratio=1 performed best for both the mesoscale and synoptic-scale experiments. This means that in both cases the model performed best when only the L1 penalty was applied, corresponding to a Lasso regression. 

Having selected the optimal hyperparameters, we fit our two models on the full training dataset dev2 using scikit-learn, using a maximum of 100,000 iterations.

\subsubsection{Fractal cascade model}

An alternative approach to stochastic downscaling is the fractal cascade approach. This was initiated by the work of \cite{lovejoy1985fractal} in the context of modelling rainfall, and is based on the assumption of a simple power scaling law connecting the larger and smaller scales. By using this law to extend the Fourier spectrum of the input field, it is possible to generate fields having appropriate small-scale structure. In this work, we use the RainFARM \citep{rebora2006rainfarm} algorithm as implemented in the PySTEPS \citep{2019pysteps} package, which is based on these principles.

In order to apply the RainFARM algorithm to our approximately Gaussian distributed data, we need to first transform the data to more closely match the distribution of rainfall. We do this by first standardising the input field by subtracting its mean and dividing by the standard deviation, and then exponentiate the standardised field:

$$X' = \mathrm{exp}\bigg(\frac{X - \mathrm{mean}(X)}{\mathrm{std}(X)}\bigg).$$

\begin{figure}[ht!]
\begin{center}
\includegraphics[scale=0.75]{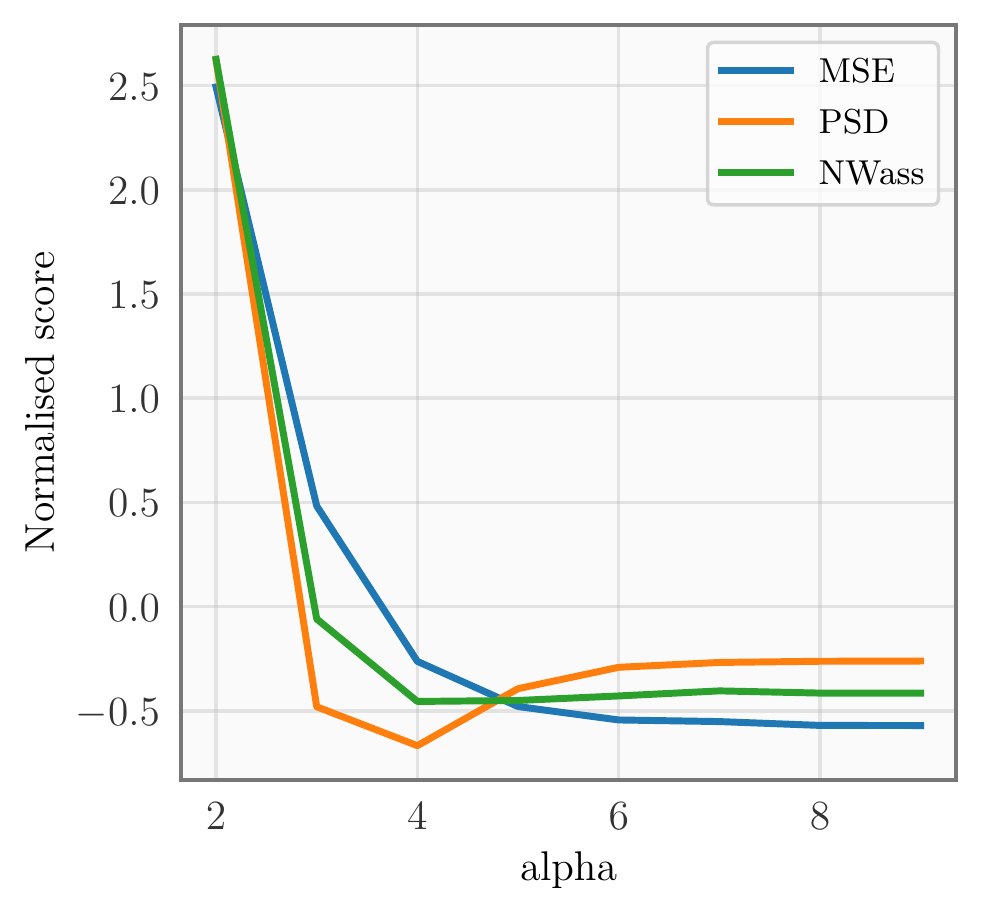}
\end{center}
\caption{Normalised training set scores for integral values of alpha for South-East England region.} \label{rainfarm-opt}
\end{figure}

The RainFARM algorithm uses a parameter called alpha, defining the slope used to extend the power spectrum (specifically, it defines the slop of the log power spectrum). There is an option to determine this parameter directly from the input field; however, we found this method gave sub-optimal results on our dataset. Instead, we fit the alpha parameter by optimising results for the power spectral score over the full training set dev2. Normalised training set scores for integral values of alpha are shown in Figure \ref{rainfarm-opt}; higher values of alpha lead to improved MSE scores in general, while the PSD and neighbourhood Wasserstein scores are minimal around alpha=4. Optimising the PSD score using Scikit-learn gave an optimal value for alpha of 3.54 over the South-East England region, 3.49 for the Midlands, and 3.50 for the Ireland region, in the synoptic experiment. For the mesoscale experiment, the corresponding values are 3.49, 3.53, and 3.51.

\subsubsection{Multiplicative cascade model}

Another kind of cascade model, also developed for rainfall disaggregation, is known as a multiplicative cascade. Instead of working with the Fourier spectrum, this approach uses stochastic weights to redistribute aggregated rainfall values among the four component sub-grid cells at the next resolution, repeating this process until the desired resolution is reached. The idea was originally explored by \cite{perica_model_1996}, who treated it as a form of fractal cascade by using a power scaling law to determine the parameters for the distributions generating the stochastic weights. However, multiplicative cascade models need not be fractal cascades. For our implementation, we follow the  classical equal-volume area model described in \cite{2020cascade}, which uses an empirical approach to fit the disaggregation parameters.

Since our task requires three levels of disaggregation the model requires six parameters, corresponding to the horizontal and vertical variability at each level. We follow \cite{2020cascade} in using a symmetric logit-normal distribution to model the weights, leaving the standard deviations to be estimated from the training data. In the original model, \cite{2020cascade} allows the standard deviation parameters to depend on the rainfall intensity as a polynomial. We follow the same approach, using a polynomial regression to determine the standard deviations as a function of the aggregate field value at the source level. We allow a maximal polynomial degree of five, and use a hard cut-off at zero to prevent nonsensical negative standard deviation values.

\subsection{Results}

We compare these GRF models with our five benchmarks using the verification scores described in Section \ref{verification} and the mean squared error. The results aggregated over the three regions are shown in Table \ref{results}.

Considering first the CRPS score, our GRF-T model has consistently the best performance for both the mesoscale and synoptic experiments. The other GRF models also perform well, out-performing the cascade models; the multiplicative cascade has the weakest performance overall. 

For the MSE score, the GRF-T mean scores consistently best, with bicubic interpolation and ElasticNet regression also performing well. As expected, the stochastic models have generally higher MSE scores than the deterministic models, particularly the multiplicative cascade model.

The comparison is less clear for the neighbourhood Wasserstein and PSD scores, with the relative performance varying depending on the experiment scale. The RainFARM model performs well according to the PSD score, outperforming the other models in the synoptic experiment. The GRF-T model also performs well, as does the original low resolution input. For the neighbourhood Wasserstein score, the GRF-T model performs well, as does the GRF-T mean, with the mean performing better in the mesoscale experiment and the samples in the synoptic experiment.   

\begin{figure}[ht!]
\begin{center}
\includegraphics[width=0.8\columnwidth]{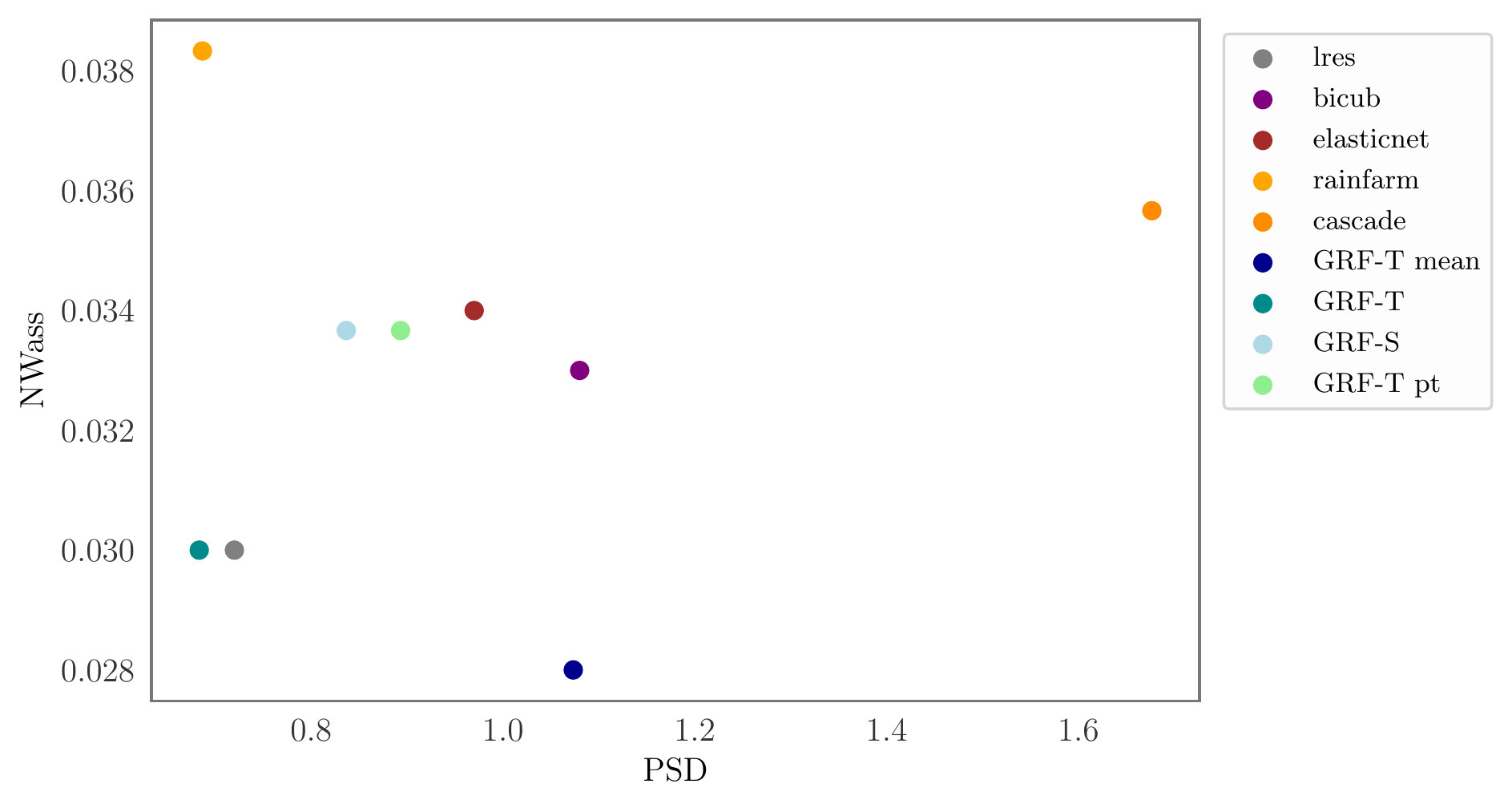}
\end{center}
\caption{Scatter plot of PSD and neighbourhood Wasserstein scores for mesoscale experiment (aggregate over all locations).} \label{scatter4}
\end{figure}

\begin{figure}[ht!]
\begin{center}
\includegraphics[width=0.8\columnwidth]{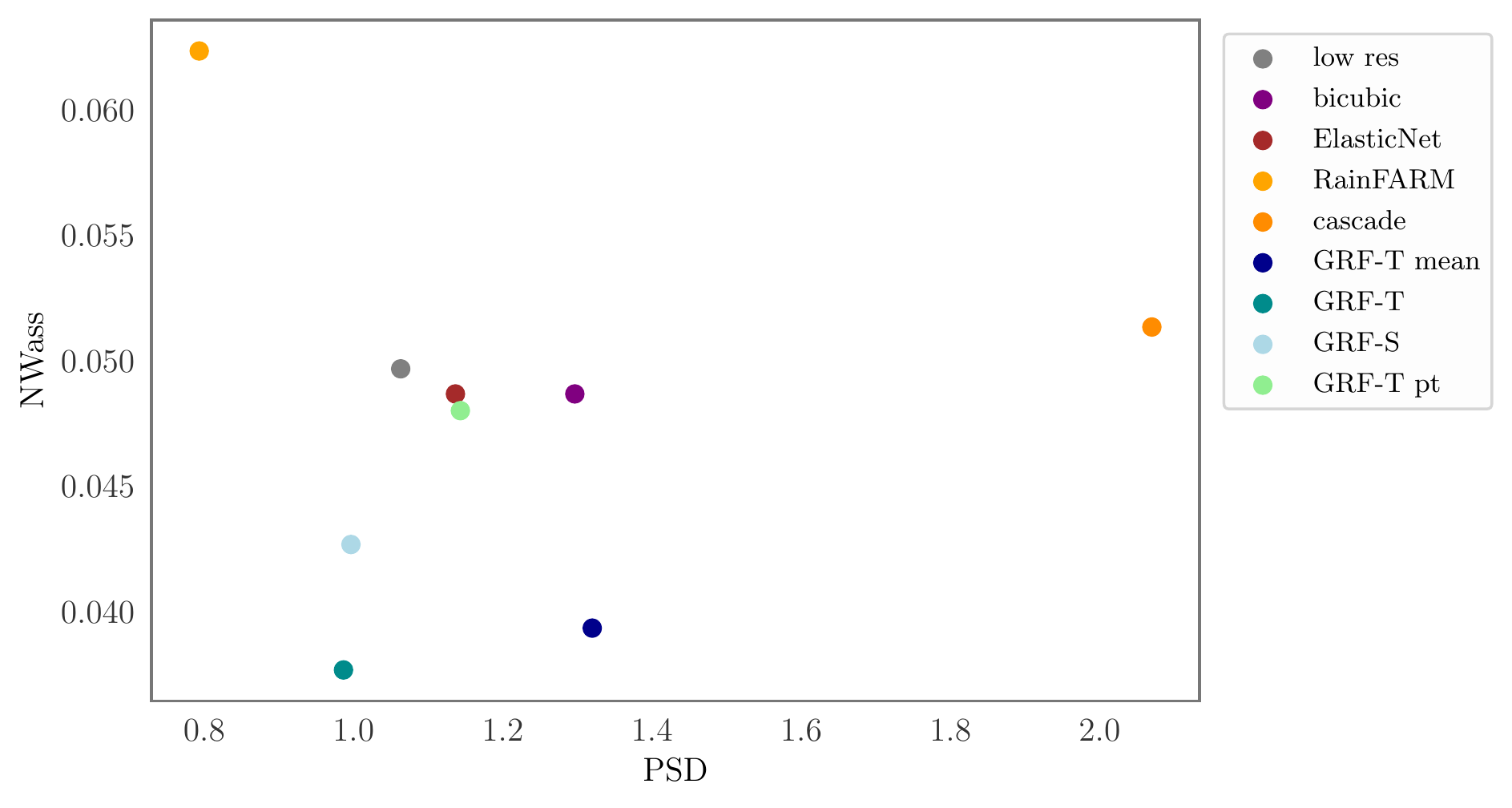}
\end{center}
\caption{Scatter plot of PSD and neighbourhood Wasserstein scores for synoptic experiment (aggregate over all locations).} \label{scatter8}
\end{figure}

A comparison of the model performance on both the neighbourhood Wasserstein and PSD scores together gives a clearer picture (Figures \ref{scatter4} and \ref{scatter8}). The models are plotted against the two scores on perpendicular axes, so that models which perform better on both scores  appear towards the bottom left corner. The GRF-T model is near to the bottom left in both the synoptic and mesoscale experiments. In the mesoscale case, the GRF-T model and GRF-T mean form a Pareto front, with the GRF-T mean scoring best for the neighbourhood Wasserstein score but less well for the PSD score. On the other hand, for the synoptic experiment the GRF-T and RainFARM models form a Pareto front, with RainFARM scoring best for the PSD score but worst for the neighbourhood Wasserstein score.

\begin{figure}[ht!]
\begin{center}
\includegraphics[width=0.8\columnwidth]{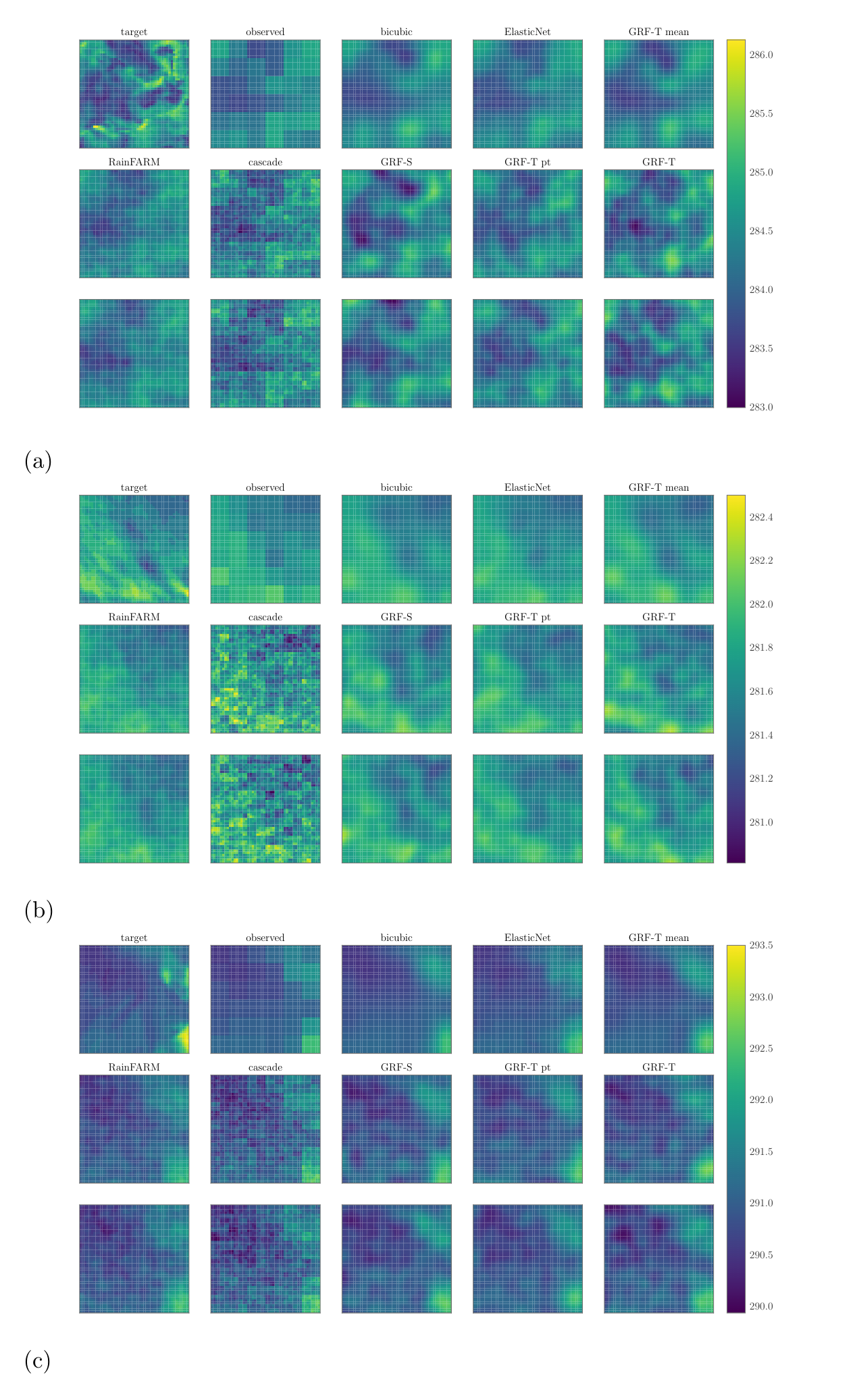}
\end{center}
\caption{Example outputs for the synoptic GRF-T model and benchmarks for South East England region. Subfigures show (a) best, (b) median and (c) worst performance of GRF-T compared to bicubic interpolation in terms of neighbourhood Wasserstein score. Area shown is approximately $106 \ \mathrm{km}^2.$}
\label{sample_output_se_8}
\end{figure}

\begin{figure}[ht!]
\begin{center}
\includegraphics[width=0.8\columnwidth]{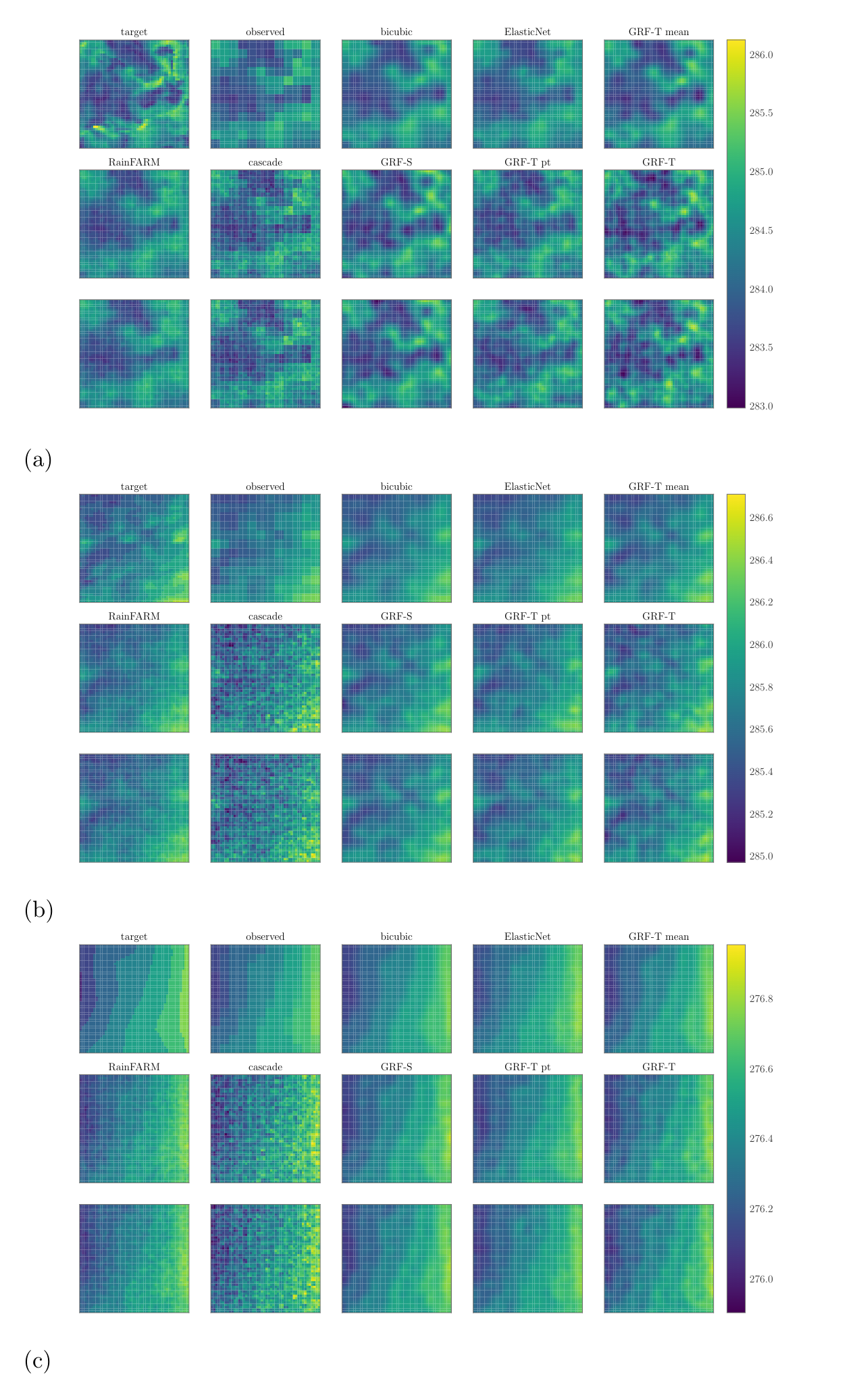}
\end{center}
\caption{Example outputs for the mesoscale GRF-T model and benchmarks for South East England region. Subfigures show (a) best, (b) median and (c) worst performance of GRF-T compared to bicubic interpolation in terms of neighbourhood Wasserstein score. Area shown is approximately $106 \ \mathrm{km}^2.$}
\label{sample_output_se_4}
\end{figure}

\begin{figure}[ht!]
\begin{center}
\includegraphics[width=0.8\columnwidth]{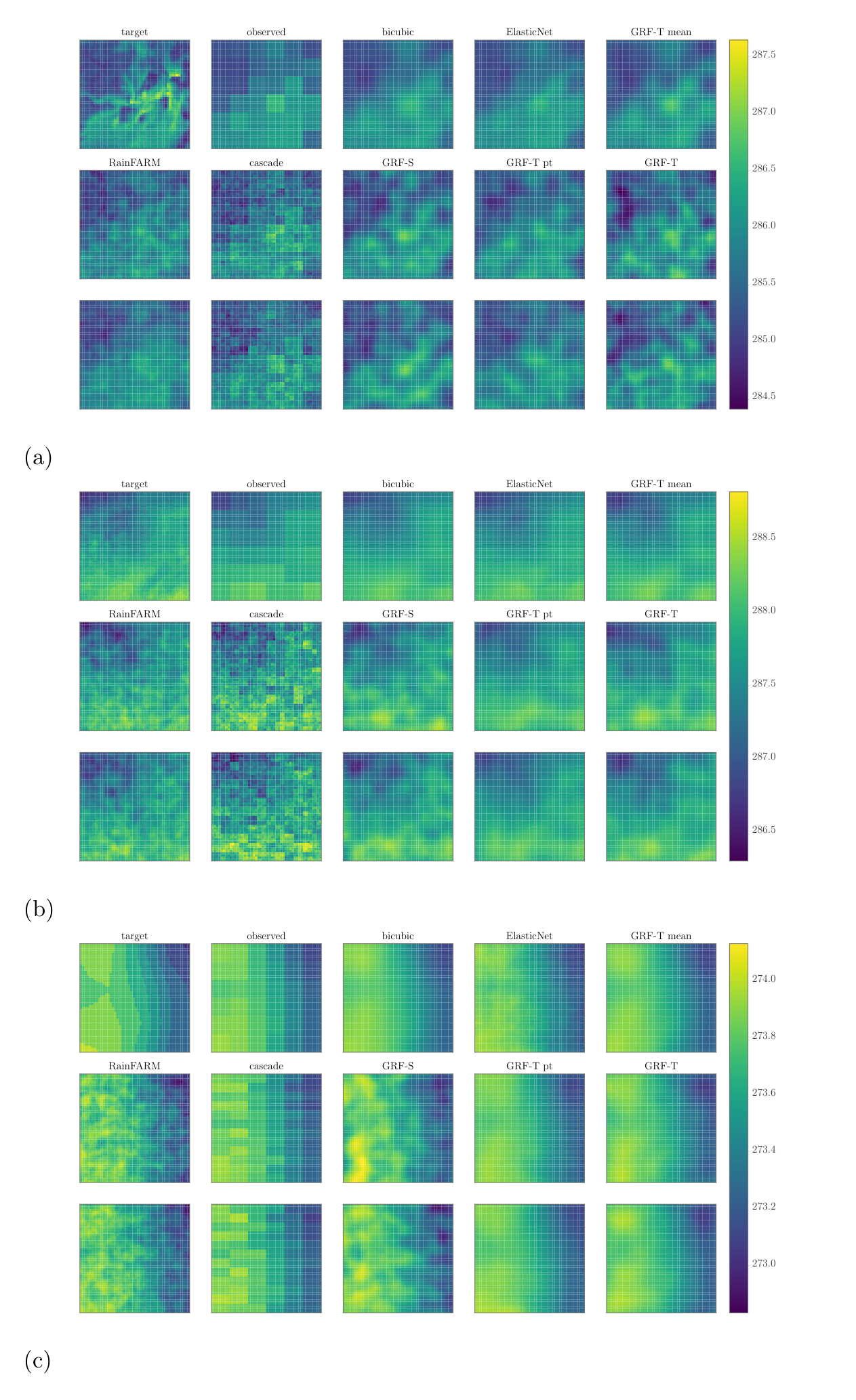}
\end{center}
 \caption{Example outputs for the synoptic GRF-T model and benchmarks for Midlands region. Subfigures show (a) best, (b) median and (c) worst performance of GRF-T compared to bicubic interpolation in terms of neighbourhood Wasserstein score. Area shown is approximately $106 \ \mathrm{km}^2.$}
\label{sample_output_mid_8}
\end{figure}

\begin{figure}[ht!]
\begin{center}
\includegraphics[width=0.8\columnwidth]{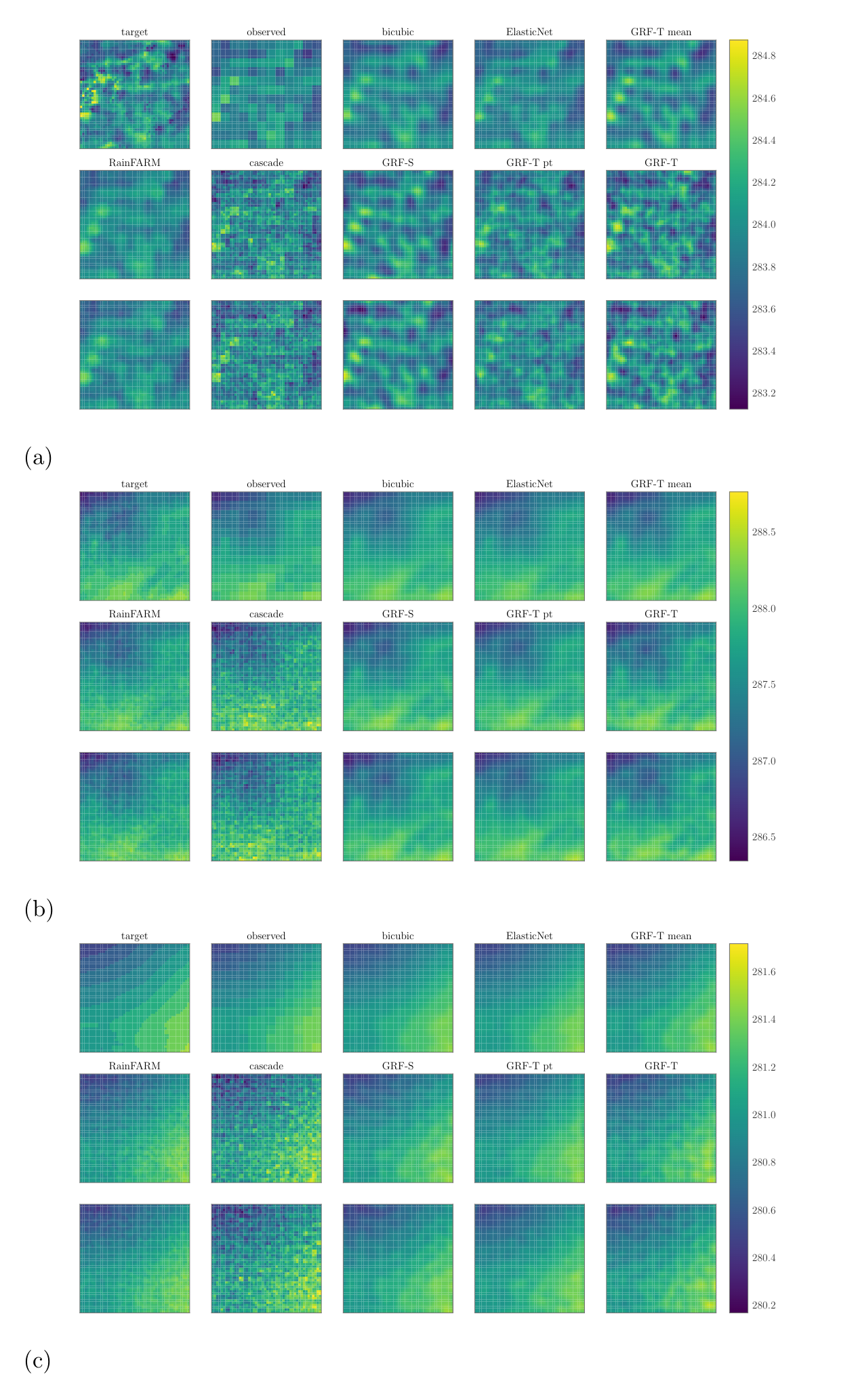}
\end{center}
 \caption{Example outputs for the mesoscale GRF-T model and benchmarks for Midlands region. Subfigures show (a) best, (b) median and (c) worst performance of GRF-T compared to bicubic interpolation in terms of neighbourhood Wasserstein score. Area shown is approximately $106 \ \mathrm{km}^2.$}
\label{sample_output_mid_4}
\end{figure}

\begin{figure}[ht!]
\begin{center}
\includegraphics[width=0.8\columnwidth]{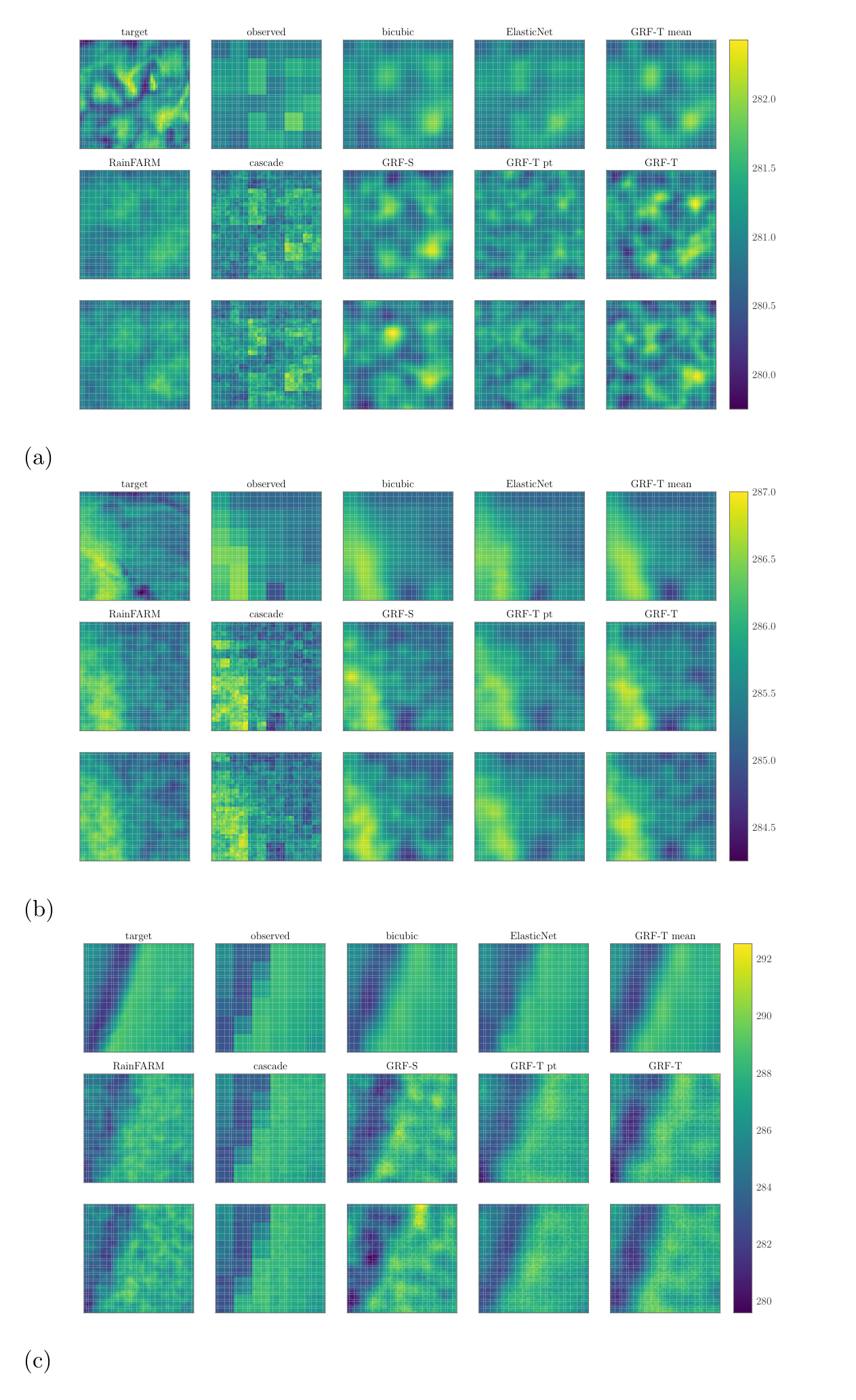}
\end{center}
\caption{Example outputs for the synoptic GRF-T model and benchmarks for Ireland region. Subfigures show (a) best, (b) median and (c) worst performance of GRF-T compared to bicubic interpolation in terms of neighbourhood Wasserstein score. Area shown is approximately $106 \ \mathrm{km}^2.$}
\label{sample_output_ei_8}
\end{figure}

\begin{figure}[ht!]
\begin{center}
\includegraphics[width=0.8\columnwidth]{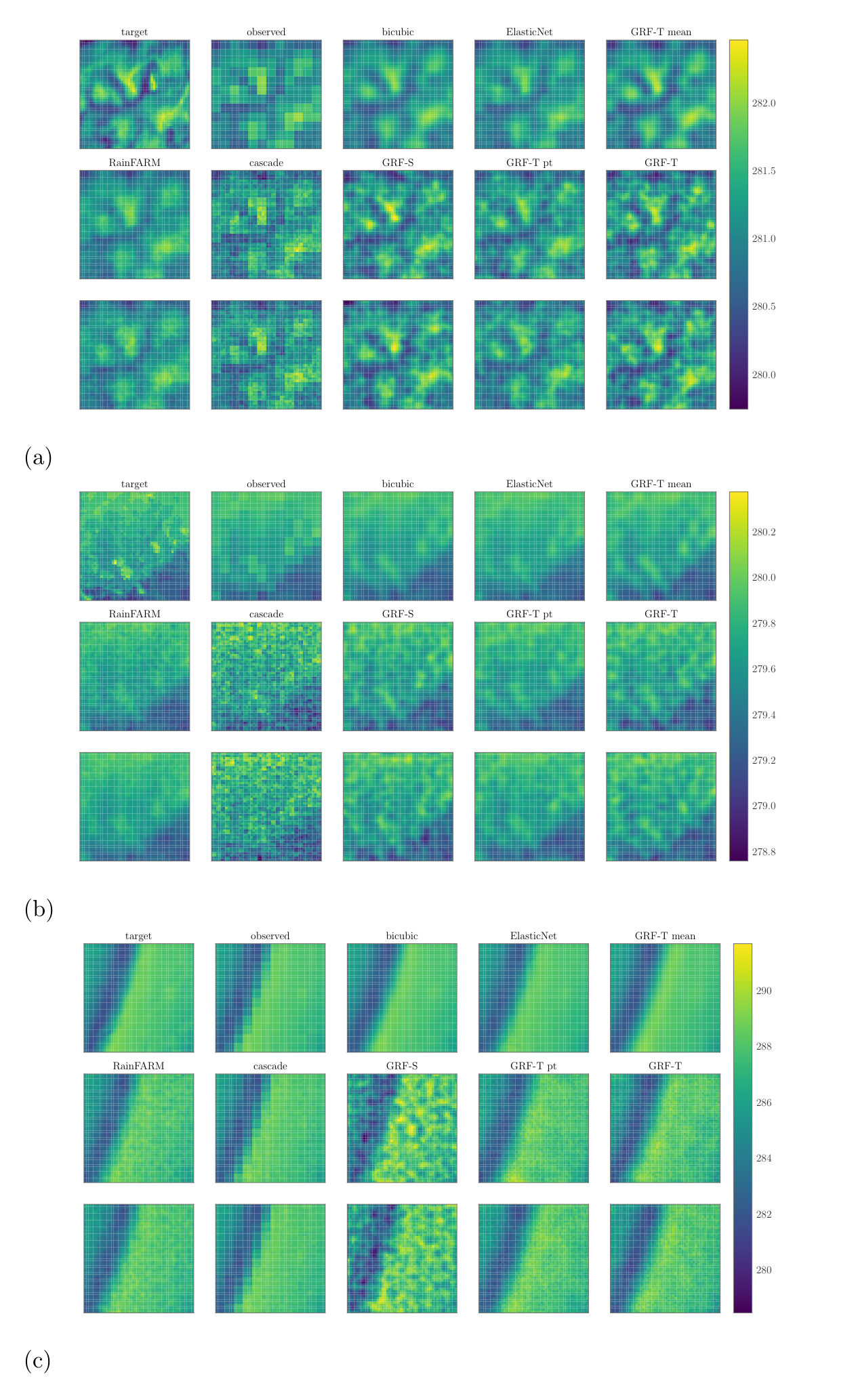}
\end{center}
\caption{Example outputs for the mesoscale GRF-T model and benchmarks for Ireland region. Subfigures show (a) best, (b) median and (c) worst performance of GRF-T compared to bicubic interpolation in terms of neighbourhood Wasserstein score. Area shown is approximately $106 \ \mathrm{km}^2.$}
\label{sample_output_ei_4}
\end{figure}

Sample outputs for the models and benchmarks can be found in Figures \ref{sample_output_se_8}, \ref{sample_output_se_4}, \ref{sample_output_mid_8}, \ref{sample_output_mid_4}, \ref{sample_output_ei_8} and \ref{sample_output_ei_4}. The subfigures show the samples in which the GRF samples perform best, median and worst compared with bicubic interpolation with respect to the neighbourhood Wasserstein score. In the best cases, the GRF-T model introduces variability which is missing from the bicubic interpolation, as well as better capturing the predictable component where it extends beyond the value range of the coarse-grained observation. In the worst cases, the GRF-T model can introduce spurious variability (e.g. Figure \ref{sample_output_mid_4}c).

\begin{table}
\begin{center}
 \begin{tabular}{|| c || c || c | c | c | c ||} 
 \hline\hline
 convolution & model & MSE & PSD Wass & Nd Wass (4) & CRPS \\ [0.5ex] 
 \hline\hline
 $4\times4$ & lres & 0.017 & 0.72 & 0.030 & - \\ 
 \hline
 & bicub & 0.012 & 1.08 & 0.033 & - \\ 
 \hline
 & elasticnet & 0.012 & 0.97 & 0.034 & - \\
 \hline
 & rainfarm & 0.019 & 0.69 & 0.038 & 0.056 \\
 \hline
 & cascade & 0.035 & 1.68 & 0.036 & 0.066 \\
 \hline
 & GRF-S samples & 0.032 & 0.84 & 0.034 & 0.051 \\
 \hline
 & GRF-T-pt samples & 0.018 & 0.89 & 0.034 & 0.053 \\ 
 \hline
 & GRF-T mean & \textbf{0.009} & 1.07 & \textbf{0.028} & - \\ 
 \hline
 & GRF-T samples & 0.018 & \textbf{0.68} & 0.030 & \textbf{0.046} \\
 \hline\hline\\
 \hline\hline
 $8\times8$ & lres & 0.040 & 1.06 & 0.050 & - \\ 
 \hline
 & bicub & 0.029 & 1.30 & 0.049 & - \\ 
 \hline
 & elasticnet & 0.030 & 1.14 & 0.049 & - \\
 \hline
 & rainfarm & 0.059 & \textbf{0.79} & 0.062 & 0.092 \\
 \hline
 & cascade & 0.078 & 2.07 & 0.051 & 0.102 \\
 \hline
  & GRF-S samples & 0.055 & 1.00 & 0.043 & 0.080 \\
 \hline
 & GRF-T-pt samples & 0.039 & 1.14 & 0.048 & 0.082 \\ 
 \hline
 & GRF-T mean & \textbf{0.025} & 1.32 & 0.039 & - \\ 
 \hline
 & GRF-T samples & 0.041 & 0.99 & \textbf{0.038} & \textbf{0.076} \\ 
 \hline
\end{tabular}
\end{center}
\caption{Results for aggregated over all three regions. Deterministic scores for GRF samples report an average of 20 samples. Neighbourhood Wasserstein score is reported for a square neighbourhood of size-length four.}
\label{results}
\end{table}

%
%

%

%
%

%

%
%

%

\section{Conclusions}

We have argued that downscaling to convective scales calls for a stochastic approach due to inherent chaotic variability at these scales. We have proposed a model for variables which are approximately Gaussian distributed, based on a spatial statistical method called Gaussian random fields. An advantage of our model is that it only requires a single input field; not being data-driven, it does not require a training dataset, only the estimation of model parameters. 

We have argued that special attention needs to be paid to verification of stochastic downscaling models, as standard scores such as MSE can be misleading when considered in isolation. In this work, we have therefore considered a suite of verification scores which includes point-based, probabilistic, and spatial scores (including one which is new to the literature). This multifaceted approach to verification can greatly improve progress in probabilistic and stochastic environmental modelling.


Overall, our model compares well to the benchmarks we have considered. It has the lowest ensemble CRPS in all cases, which demonstrates that it produces skilful forecasts in a probabilistic sense. The model mean has the lowest MSE score of all the models, including the data-driven deterministic ElasticNet model. Finally, the model scores extremely well when considering the neighbourhood Wasserstein and PSD scores together (Figures \ref{scatter4} and \ref{scatter8}), although it is not uniformly the best when considering either score individually. 

In order to test the effectiveness of different aspects of our model we have compared against two alternative GRF models, one using a time-stationary covariance, and one using point conditioning instead of spatial conditioning. Our GRF-T model outperforms both alternatives, giving the best performance in the PSD, neighbourhood Wasserstein, and CRPS scores. Thus, it appears that both the spatial conditioning approach and the adaptive covariance help to improve the model performance.

With regards to future work, there are various extensions to our model which could be considered. For example, our model can be extended by considering more general methods of estimating the high-resolution covariance without any change to the conditioning step. A clear avenue for future work is therefore to replace our simple method of covariance estimation with one which can account for spatial non-stationarity. 

Going further in this direction, the isotropic assumption used in our covariance model is likely to be unjustified for convective-scale weather, since factors such as synoptic drivers and atmospheric waves can introduce a major directional component. In principle, this can be easily dealt with by considering a linear transformation of the input space defined by a positive semidefinite matrix $M$ which captures the directional dependence, so that the distance $r^2$ between points $\mathbf{x}$ and $\mathbf{x'}$ is given by $r^2(\mathbf{x}, \mathbf{x'}) = (\mathbf{x} - \mathbf{x'})^T M (\mathbf{x} - \mathbf{x'})$ \citep{Rasmussen2006Gaussian}. In addition, the present approach of estimating the covariance for each time-step independently could be replaced by one which makes use of temporal continuity, such as an autoregressive model or a Kalman filter (e.g. \cite{carron_machine_2016}). 

Besides the assumptions used for covariance estimation, a core limitation of our current method is its assumption of Gaussianity. Another branch of future work could consider extensions to non-Gaussian variables, particularly cloud cover and rainfall.


\begin{appendices}
\section{Conditioning a Gaussian density}

\begin{thm1}
If $x_o$ and $x_t$ follow a joint Gaussian distribution so that 

$$\begin{bmatrix}
x_t \\ x_o \end{bmatrix} \sim \mathcal{N} \left( \begin{bmatrix}\mu_t \\ \mu_o \end{bmatrix}, \begin{bmatrix}C_{t} & C_{o,t}\\
C_{t,o} & C_{o} \end{bmatrix} \right)$$

\noindent then the conditional distribution $p(x_t | x_o)$ is Gaussian with mean $\mu_t - C_{o,t}C_o^{-1}(x_o - \mu_o)$ and covariance $C_t - C_{o,t}C_o^{-1}C_{t,o}$.
\end{thm1}
\begin{proof}
\label{cond_proof}
Our exposition here follows \cite{do_more_2008}. We have the conditional density formula
\begin{equation}
\begin{aligned}
p(x_t|x_o) = \frac{p(x_t,x_o)}{p(x_o)} &= \frac{1}{Z} \  \mathrm{exp}\left(-\frac{1}{2} 
\begin{bmatrix}x_t - \mathbf{\mu}_t \\ x_o - \mathbf{\mu}_o \end{bmatrix}^{T}
\begin{bmatrix}C_t & C_{o,t} \\ C_{t,o} & C_o \end{bmatrix}^{-1}
\begin{bmatrix}x_t - \mathbf{\mu}_t \\ x_o - \mathbf{\mu}_o \end{bmatrix} \right) \\
&= \frac{1}{Z} \  \mathrm{exp}\left(-\frac{1}{2} 
\begin{bmatrix}x_t - \mathbf{\mu}_t \\ x_o - \mathbf{\mu}_o \end{bmatrix}^{T}
\begin{bmatrix}P_t & P_{o,t} \\ P_{t,o} & P_o \end{bmatrix}
\begin{bmatrix}x_t - \mathbf{\mu}_t \\ x_o - \mathbf{\mu}_o \end{bmatrix} \right)
\end{aligned}
\end{equation}
\noindent where we have extracted all terms not depending on $x_t$ into a normalization constant $Z$ and used the precision matrix notation for the inverse covariance, $P = C^{-1}$.

Expanding the matrix product gives us

\begin{equation}
\begin{aligned}
    p(x_t|x_o) = \frac{1}{Z}\ \mathrm{exp}\bigg(-\frac{1}{2}
    \big[&(x_t - \mathbf{\mu}_t)^T P_t (x_t - \mathbf{\mu}_t) + 
    (x_o - \mathbf{\mu}_o)^T P_{t,o} (x_t - \mathbf{\mu}_t) \\ &+
    (x_t - \mathbf{\mu}_t)^T P_{o,t} (x_o - \mathbf{\mu}_o) +
    (x_o - \mathbf{\mu}_o)^T P_o (x_o - \mathbf{\mu}_o) \big] \bigg).
\end{aligned}
\end{equation}

Now we can complete the square to give a quadratic term in $x_t$ plus a constant term. We use the formula

\begin{equation}
z^T A z + 2 z^T b + c = \big(z + A^{-1}b\big)^T A \big(z + A^{-1}b\big) + c - b^T A^{-1} b
\end{equation}

\noindent for a symmetric matrix $A$, and substitute

\begin{equation}
\begin{aligned}
    z &= (x_t - \mu_t) \\
    A &= P_t \\
    b &= P_{o,t}(x_o - \mu_o) \\
    c &= (x_o - \mu_o)^T P_o (x_o - \mu_o)
\end{aligned}
\end{equation}

\noindent to give

\begin{equation}
\begin{aligned}
p(x_t|x_o) = \frac{1}{Z}\ \mathrm{exp}\bigg(-\frac{1}{2}
\big[&\big(x_t - \mu_t + P_t^{-1} P_{o,t}(x_o - \mu_o)\big)^T P_t \big(x_t - \mu_t + P_t^{-1} P_{o,t}(x_o - \mu_o)\big) \\ &+
(x_o - \mu_o)^T P_o (x_o - \mu_o) - (x_o - \mu_o)^T P_{t,o} P_t^{-1} P_{o,t} (x_o -\mu_o) \big]\bigg)
\end{aligned}
\end{equation}


We can then move the constant term not depending on $x_t$ out of the exponential into the normalization term:

\begin{equation}
\begin{aligned}
p(x_t|x_o) = \frac{1}{Z'}\ \mathrm{exp}\bigg(-\frac{1}{2}
\big(x_t - \mu_t + P_t^{-1} P_{o,t}(x_o - \mu_o)\big)^T P_t \big(x_t - \mu_t + P_t^{-1} P_{o,t}(x_o - \mu_o)\big)\bigg)
\end{aligned}
\end{equation}

This gives a Gaussian density in terms of the precision $P$, with mean $\mu_t - P_t^{-1}P_{o,t}(x_o - \mu_o)$ and covariance $P_t^{-1}$. Now we would like to put this in terms of the covariance $C$. We can do so using the matrix inversion formula by noting that

\begin{equation}
\begin{bmatrix}C_t & C_{o,t} \\ C_{t,o} & C_o \end{bmatrix} = 
\begin{bmatrix}(P_t - P_{o,t}P_o^{-1}P_{t,o})^{-1} & 
-P_t^{-1}P_{o,t}(P_o - P_{t,o}P_t^{-1}P_{o,t})^{-1} \\ 
-(P_o - P_{t,o}P_t^{-1}P_{o,t})^{-1}P_{t,o}P_t & 
(P_o - P_{t,o}P_t^{-1}P_{o,t})^{-1} \end{bmatrix}
\end{equation}
\noindent and so

\begin{equation}
    \mu_{t|o} = \mu_t - P_t^{-1}P_{o,t}(x_o - \mu_o) = \mu_t - C_{o,t}C_o^{-1}(x_o - \mu_o).
\end{equation}

\noindent On the other hand, we have

\begin{equation}
\begin{bmatrix}P_t & P_{o,t} \\ P_{t,o} & P_o \end{bmatrix} = 
\begin{bmatrix}(C_t - C_{o,t}C_o^{-1}C_{t,o})^{-1} & 
-C_t^{-1}C_{o,t}(C_o - C_{t,o}C_t^{-1}C_{o,t})^{-1} \\ 
-(C_o - C_{t,o}C_t^{-1}C_{o,t})^{-1}C_{t,o}C_t & 
(C_o - C_{t,o}C_t^{-1}C_{o,t})^{-1} \end{bmatrix}
\end{equation}
\noindent which gives us

\begin{equation}
    C_{t|o} = P_t^{-1} = C_t - C_{o,t}C_o^{-1}C_{t,o}
\end{equation}

\noindent as required.

\end{proof}

\begin{thm1}
For a continuous Gaussian process with covariance kernel $k$, the conditional density of a point value $x_t$ conditioned on a spatial average $x_o = \frac{1}{\int_{T} 1 d\tau} \int_{T} \,x_{\tau} d\tau$ is given by 
\begin{equation}
\begin{aligned}
    \mu_{t|o} =  C_{o,t}C_o^{-1}(x_o - \mu_o)\\
    C_{t|o} = C_t - C_{o,t}C_o^{-1}C_{t,o}
\end{aligned}
\end{equation}
with
\begin{equation}
\begin{aligned}
    C_{(t)i,j} &= k(x_i, x_j)\\
    C_{(o,t)i,j} = C_{(t,o)j,i} &= \int_{T_j} k(x_i, x_{\tau_j}) \ d\tau_j\\
    C_{(o)i,j} &= \int_{T_i} \int_{T_j} k(x_{\tau_i}, x_{\tau_j}) \ d\tau_i \ d\tau_j
\end{aligned}
\end{equation}

\noindent and
\begin{equation}
    \mu_o = \mathbb{E}(x_o) = \int_{T} \mathbb{E}(x_{\tau}) d\tau.
\end{equation}

\end{thm1}
\label{cond_proof_cont}

\begin{proof}
Conditioning on a spatial average is closely related to conditioning on an integral of the underlying process. In fact, the average is equivalent to an integral multiplied by a scaling factor  $\frac{1}{\int_{T} 1 d\tau} \int_{T} \,x_{\tau} d\tau$. For the remainder of this section, we take  $I = \int_{T} \,x_{\tau} d\tau$ to be such a rescaled integral so that  $\int_{T} 1 d\tau = 1$. 

In what follows, we will need to make reference to the space of samples $\mathcal{H}$ of the Gaussian process. We will take $s \in \mathcal{H}$ to be any possible sample of the process, and $p(s)$ to be the probability density of the sample $s$.

Considering first the mean, we can use this terminology to write the expected value of the observed spatial integral $I$ as
\begin{equation}
\label{EI}
	\mathbb{E}(I) = \int_{\mathcal{H}} I(s) p(s) ds
\end{equation}

\noindent where $I(s)$ is the integral $I$ taken over a sample $s$. Here, we have rewritten the expectation as a weighted integral over all possible samples of our process.

We can rewrite this equation by noting that the probability density of process samples is symmetrically weighted around the mean, that is:
\begin{equation}
	p(\mathbb{E}(s) + r) = p(\mathbb{E}(s) - r).
\end{equation}

\noindent On the other hand, the value of the spatial integral $I$ follows the relation:
\begin{equation}
\begin{aligned}
	I(\mathbb{E}(s) + r) &= I(\mathbb{E}(s)) + I(r) \\
	I(\mathbb{E}(s) - r) &= I(\mathbb{E}(s)) - I(r).
\end{aligned}
\end{equation}

\noindent When we consider Equation \ref{EI} the opposite-signed terms involving $r$ then cancel, leaving
\begin{equation}
\begin{aligned}
	\mathbb{E}(I) &= \int_{\mathcal{H}} I(s) p(s) ds \\
	                      &= \int_{\mathcal{H}} I(\mathbb{E}(s)) p(s) ds \\
	                      &= I(\mathbb{E}(s)) \int_{\mathcal{H}}  p(s) ds \\
	                      &= I(\mathbb{E}(s)).
\end{aligned}
\end{equation}

For the covariance, we need to consider both the covariance between an integral and a point, and the covariance between two integrals (the covariance between two points being the standard form). We begin with the covariance between a point $x$ and an integral $I = \int_{T} \,x_{\tau} d\tau$. We have
\begin{equation}
\begin{aligned}
	\mathrm{cov}(x, I) = \mathbb{E}(xI) - \mathbb{E}(x) \mathbb{E}(I).
\end{aligned}
\end{equation}

\noindent Considering the term $\mathbb{E}(xI)$ we have
\begin{equation}
\begin{aligned}
	\mathbb{E}(xI) &= \mathbb{E}\bigg(x \int_{T} x_{\tau} \ d\tau \bigg) \\
		                &= \mathbb{E}\bigg(\int_{T} x x_{\tau} \ d\tau \bigg) \\
		                &= \int_{\mathcal{H}} \int_{T} x x_{\tau} \ d\tau \ p(s) \ ds \\
		                &= \int_{\mathcal{H}} \int_{T} x x_{\tau} \ p(s) \ d\tau \ ds.
\end{aligned}
\end{equation}

Now, $x x_{\tau}$ is continuous everywhere in $\mathcal{H} \times T$ where $s$ is continuous. But if the process we are considering uses a standard kernel function such as the Matern or squared exponential kernel, its sample paths are always continuous \citep{paciorek_cj_nonstationary_2003}. Thus the integrand is a continuous function and we can change the order of integration to give
\begin{equation}
\begin{aligned}
	\mathbb{E}(xI) &= \int_{\mathcal{H}} \int_{T} x x_{\tau} \ p(s) \ d\tau \ ds \\
			       &= \int_{T} \int_{\mathcal{H}} x x_{\tau} \ p(s) \ ds \ d\tau \\
			       &= \int_{T} \mathbb{E}(x x_{\tau}) \ d\tau \\
			       &= \int_{T} \mathrm{cov}(x, x_{\tau}) + \mathbb{E}(x)\mathbb{E}(x_{\tau}) \ d\tau \\
			       &= \int_{T} \mathrm{cov}(x, x_{\tau}) \ d\tau + \mathbb{E}(x)\mathbb{E}(I)
\end{aligned}
\label{arg}
\end{equation}

\noindent by eqn \ref{EI}, and so
\begin{equation}
\begin{aligned}
	\mathrm{cov}(x, I)  &= \int_{T} \mathrm{cov}(x, x_{\tau}) \ d\tau + \mathbb{E}(x)\mathbb{E}(I) - 
	\mathbb{E}(x)\mathbb{E}(I)\\
				     &= \int_{T} \mathrm{cov}(x, x_{\tau}) \ d\tau \\
			             &= \int_{T} k(x, x_{\tau}) \ d\tau
\end{aligned}
\end{equation}

\noindent for covariance kernel $k$.

We now consider the covariance between two integrals, $I_1 = \int_{T_1} \,x_{{\tau}_1} d\tau_1$ and $I_2 = \int_{T_2} \,x_{\tau_2} d\tau_2$. As before, we have
\begin{equation}
\begin{aligned}
	\mathrm{cov}(I_1, I_2) = \mathbb{E}(I_1 I_2) - \mathbb{E}(I_1) \mathbb{E}(I_2).
\end{aligned}
\end{equation}

\noindent and considering the first term gives us
\begin{equation}
\begin{aligned}
	\mathbb{E}(I_1 I_2) = \mathbb{E} \bigg( \int_{T_1} x_{\tau_1} d\tau_1  \cdot \int_{T_2} x_{\tau_2} d\tau_2 \bigg).
\end{aligned}
\end{equation}

\noindent Since the domains of integration are independent and the integrands are continuous by the argument above, we can rewrite this as a double integral
\begin{equation}
\begin{aligned}
	\mathbb{E}(I_1 I_2) = \mathbb{E} \bigg( \int_{T_1} \int_{T_2} x_{\tau_1} x_{\tau_2} \ d\tau_1 \ d\tau_2 \bigg).
\end{aligned}
\end{equation}

\noindent Now the argument of \ref{arg} gives 
\begin{equation}
\begin{aligned}
	\mathbb{E}(I_1 I_2) &= \int_{T_1} \int_{T_2} \mathbb{E} (x_{\tau_1} x_{\tau_2}) \ d\tau_1 \ d\tau_2 \\
				       &= \int_{T_1} \int_{T_2} \mathrm{cov} (x_{\tau_1}, x_{\tau_2}) + \mathbb{E}(x_{\tau_1}) \mathbb{E}(x_{\tau_2}) \ d\tau_1 \ d\tau_2 \\
				       &= \int_{T_1} \int_{T_2} \mathrm{cov} (x_{\tau_1}, x_{\tau_2})  \ d\tau_1 \ d\tau_2 + \mathbb{E}(I_1) \mathbb{E}(I_2) \\
\end{aligned}
\end{equation}

\noindent so that
\begin{equation}
\begin{aligned}
	\mathrm{cov}(I_1, I_2)  &= \int_{T_1} \int_{T_2} \mathrm{cov}(x_{\tau_1}, x_{\tau_2}) \ d\tau_1 \ d\tau_2\\
			                     &= \int_{T_1} \int_{T_2} k(x_{\tau_1}, x_{\tau_2}) \ d\tau_1 \ d\tau_2
\end{aligned}
\end{equation}

\noindent for covariance kernel $k$.

We can combine this with our equations from the previous section, namely
\begin{equation}
\begin{aligned}
    \mu_{t|o} =  C_{o,t}C_o^{-1}(x_o - \mu_o)\\
    C_{t|o} = C_t - C_{o,t}C_o^{-1}C_{t,o}
\end{aligned}
\end{equation}

\noindent with integrals $I = \int_{T} \,x_{\tau} d\tau$ playing the role of the observed variables $x_o$, and target variables $x_t$ remaining discrete as before. We now have

\begin{equation}
\begin{aligned}
    C_{(t)i,j} &= k(x_i, x_j)\\
    C_{(o,t)i,j} = C_{(t,o)j,i} &= \int_{T_j} k(x_i, x_{\tau_j}) \ d\tau_j\\
    C_{(o)i,j} &= \int_{T_i} \int_{T_j} k(x_{\tau_i}, x_{\tau_j}) \ d\tau_i \ d\tau_j
\end{aligned}
\end{equation}

\noindent and
\begin{equation}
    \mu_o = \mathbb{E}(x_o) = \int_{T} \mathbb{E}(x_{\tau}) dt.
\end{equation}

\end{proof}

\noindent In the discrete case, this corresponds to the relations in equation \ref{eq:cond}.


\section{Computation of 1-Wasserstein distance}
\label{wass_appx}
By using the earth movers's interpretation of the 1-Wasserstein distance, we know that it can be found by summing the distance moved by each point. Each value in the first list must map uniquely onto a value of the second list. So, for any chosen ordering on the first list, we can find an ordering of the second list such that the summed difference between the ordered lists is equal to the 1-Wasserstein distance, and specifically this ordering is the one which minimises the summed difference. Since we have the freedom to choose an ordering on the first list, let's assume it to be sorted in increasing order. Suppose we have an ordering on the second list. Clearly, any pair of two points in the second list are either in increasing order or they are swapped. We will proceed by showing that exchanging two swapped points will never increase the total difference, it must either reduce the difference or leave it unchanged. Thus, ordering the second list in increasing order must give a minimal value for the summed difference of the two lists, giving the desired result.

It remains to prove that exchanging two swapped points of the second list will never increase the total summed distance. We will do so by enumerating cases. We take $A$ to be the ordered list and $B$ to define our mapping. The total difference is given by $\sum_i |a_i - b_i|$, with $i$ an index on $A$ and $B$. 

Consider switching a pair $b_i$ with $b_j$ in $B$ with $b_j < b_i$, and let $a_i$ and $a_j$ be the corresponding elements of $A$ with $a_i < a_j$. There are three possible cases we need to consider. One of the following must hold:
\begin{enumerate}
    \item $a_i \leq a_j \leq b_j \leq b_i$
    \item $a_i \leq b_j \leq a_j \leq b_i$
    \item $a_i \leq b_j \leq b_i \leq a_j$
    \item $b_j \leq b_i \leq a_i \leq a_j$
    \item $b_j \leq a_i \leq b_i \leq a_j$
    \item $b_j \leq a_i \leq a_j \leq b_i$
\end{enumerate}

Case 1 (a symmetrical argument holds for case 3):
\begin{align*}
|b_i - a_i| + |b_j - a_j| &= (b_i - a_i) + (b_j - a_j) \\
&= [(b_i - b_j) + (b_j - a_j) + (a_j - a_i)] + (b_j - a_j) \\
&= [(a_j - a_i) + (b_j - a_j)] + [(b_j - a_j) + (b_i - b_j)] \\
&= (b_j - a_i) + (b_i - a_j) \\
&= |b_j - a_i| + |b_i - a_j|
\end{align*}

Case 2 (a symmetrical argument holds for case 4):
\begin{align*}
|b_i - a_i| + |b_j - a_j| &= (b_i - a_i) + (a_j - b_j) \\
&= [(b_i - a_j) + (a_j - b_j) + (b_j - a_i)] + (a_j - b_j) \\
&\geq (b_j - a_i) + (b_i - a_j)\\
&= |b_j - a_i| + |b_i - a_j|
\end{align*}

Case 3 (a symmetrical argument holds for case 6):
\begin{align*}
|b_i - a_i| + |b_j - a_j| &= (b_i - a_i) + (a_j - b_j) \\
&= [(b_i - b_j) + (b_j - a_i)] + [(a_j - b_i) - (b_i - b_j)] \\
&\geq (b_j - a_i) + (a_j - b_i) \\
&= |b_j - a_i| + |b_i - a_j|
\end{align*}

Thus in all cases exchanging the elements $b_i$ and $b_j$ either reduces or leaves unchanged the total difference. This concludes our argument. 

\end{appendices}

\bibliographystyle{ametsoc2014}
\bibliography{references}

%

%

\end{document}